\documentclass[onecolumn]{emulateapj}
\usepackage{amsmath,amssymb}
\usepackage{graphicx,mathptm}
\usepackage{times}
\usepackage{natbib}
\usepackage{longtable}
\usepackage{multirow}
\usepackage{rotating}
\usepackage{epstopdf}
\usepackage{bm}
\usepackage{indentfirst}
\usepackage{booktabs}
\newcommand{\ud}{\mathrm{d}}
\newcommand{\be}{\begin{equation}}
\newcommand{\ee}{\end{equation}}
\newcommand{\bea}{\begin{eqnarray}}
\newcommand{\eea}{\end{eqnarray}}
\newcommand {\apgt} {\ {\raise-.5ex\hbox{$\buildrel>\over\sim$}}\ }
\newcommand {\aplt} {\ {\raise-.5ex\hbox{$\buildrel<\over\sim$}}\ }
\usepackage{threeparttable}
\usepackage{subfigure}
\usepackage{mathrsfs}
\usepackage{url}
\begin{document}
\title{Tracing Magnetic Fields by Atomic Alignment in Extended Radiation Fields}
\author{Heshou Zhang\altaffilmark{1} \& Huirong Yan\altaffilmark{2} \& Le Dong\altaffilmark{3} }
\altaffiltext{1}{Yuanpei College, Peking University, Beijing 100871, P.R.China;}
\altaffiltext{2}{Kavli Institute of Astronomy and Astrophysics, Peking University, Beijing 100871, P.R.China; hryan@pku.edu.cn}
\altaffiltext{3}{Yuanpei College, Peking University, Beijing 100871, P.R.China;}
\begin{abstract}
Tracing magnetic field is crucial as magnetic field plays an important role in many astrophysical processes. Earlier studies have demonstrated that Ground State Alignment (GSA) is an effective way to detect weak magnetic field $(1G\apgt B\apgt 10^{-15}G)$ in diffuse medium. We explore the atomic alignment in the presence of extended radiation field for both absorption line and emission line. The alignment in circumstellar medium, binary systems, discs, and the Local Interstellar Medium (LISM) are considered in order to study the alignment in the radiation field where the pumping source has a clear geometric structure. Furthermore, the method of multipole expansion is adopted to study GSA induced in the radiation field with unidentified pumping sources. We study the alignment in the dominant radiation components of general radiation field: dipole and quadrupole radiation field. We discuss the approximation of GSA in general radiation field by summing the contribution from the dipole and quadrupole radiation field. We conclude that GSA is a powerful tool to detect weak magnetic field in diffuse medium in general radiation field.
\end{abstract}
\section{Introduction}

The astrophysical magnetic field plays an essential role as it is found almost everywhere from relatively small scale system like solar wind, to considerably large scale system such as molecular clouds, galaxies, clusters of galaxies. Many astrophysical processes involve magnetic field including star formation, cosmic ray acceleration, accretion discs jets, etc. However, few techniques are available to detect magnetic field and the applicable environment of each technique is limited. Even the direction of magnetic field obtained from the same region of sky with different techniques differs substantially. The synergic use of different techniques is necessary (see \citealt{2013arXiv1302.3264Y}). Therefore, it is important to explore new promising magnetic tracers.

Ground State Alignment (GSA) has been demonstrated to be a powerful tool of studying magnetic field in radiation-dominated environments (see \citealt{YLfine,YLhyf,YLhanle,2012JQSRT.113.1409Y,2013arXiv1302.3264Y} for details). GSA is sensitive to the weak magnetic field in diffuse medium \citep[see][]{YLfine}. It is worth noting that optical pumping was firstly proposed by \citet{KASTLER-1950-234250}, and then the atomic alignment in the presence of magnetic field was studied in laboratory \citep[see][]{Hawkins:1955fv}. This effect was then applied to aligned atoms in toy models by \citet{Varshalovich:1968qc,Varshalovich:1971mw}. Later, the case of emission from an atom with idealized fine structure for a particular geometry of magnetic field and light beam was discussed in \citet{Landolfi:1986lh}.

The basic idea for GSA has been well illustrated: anisotropic radiation pumps the atoms with different probabilities from the sublevels of the ground state to the upper levels. The decay from the upper levels has the same probability to all the sublevels on the ground state. As a result, the occupations of the atoms on different sublevels of the ground state are changed. In the presence of magnetic field, the angular momenta of the atoms are then redistributed among the different sublevels owning to the fast magnetic precession. The alignment is altered according to the angle between magnetic field and radiation field $\theta_r$. This is magnetic realignment \citep[see][]{2012JQSRT.113.1409Y,2013arXiv1302.3264Y}.

Calculations for GSA with fine and hyperfine structure in astrophysical environment were provided in \citet{YLfine,YLhyf,YLhanle}. They demonstrated an exclusive feature of GSA that it reveals the 3D direction of magnetic field. \citet{2013Ap&SS.343..335S} discussed the applicability of GSA particularly in interplanetary medium.

Those previous works have already applied GSA to the detection of magnetic field in diffuse medium where the radiation field is a beam of light. However, GSA with general radiation field can be different. For example, when the medium is close to the radiation source, the pumping source cannot be treated as a point source. The spatial distribution of the pumping source directly decides the anisotropy of the radiation field, leading to different alignment from the case with a beam of light.

The structure of the paper is organized as follows. The physics for GSA with extended radiation field is discussed in \S 2. The general formulae of GSA are presented in \S 3. We first illustrate the alignment in the radiation field with identified pumping sources, including circumstellar medium in \S 4, binary system in \S 5, and the Local ISM in \S 6. The method of multipole expansion is illustrated and adopted to study the alignment in the radiation field with unidentified pumping sources in \S 7. Discussions and summary are provided in \S 8 and \S 9, respectively.

\section{Physics for GSA}

\subsection{Toy model for GSA}

\begin{figure*}
\centering
 \subfigure[]{
\includegraphics[width=0.44\columnwidth,
 height=0.22\textheight]{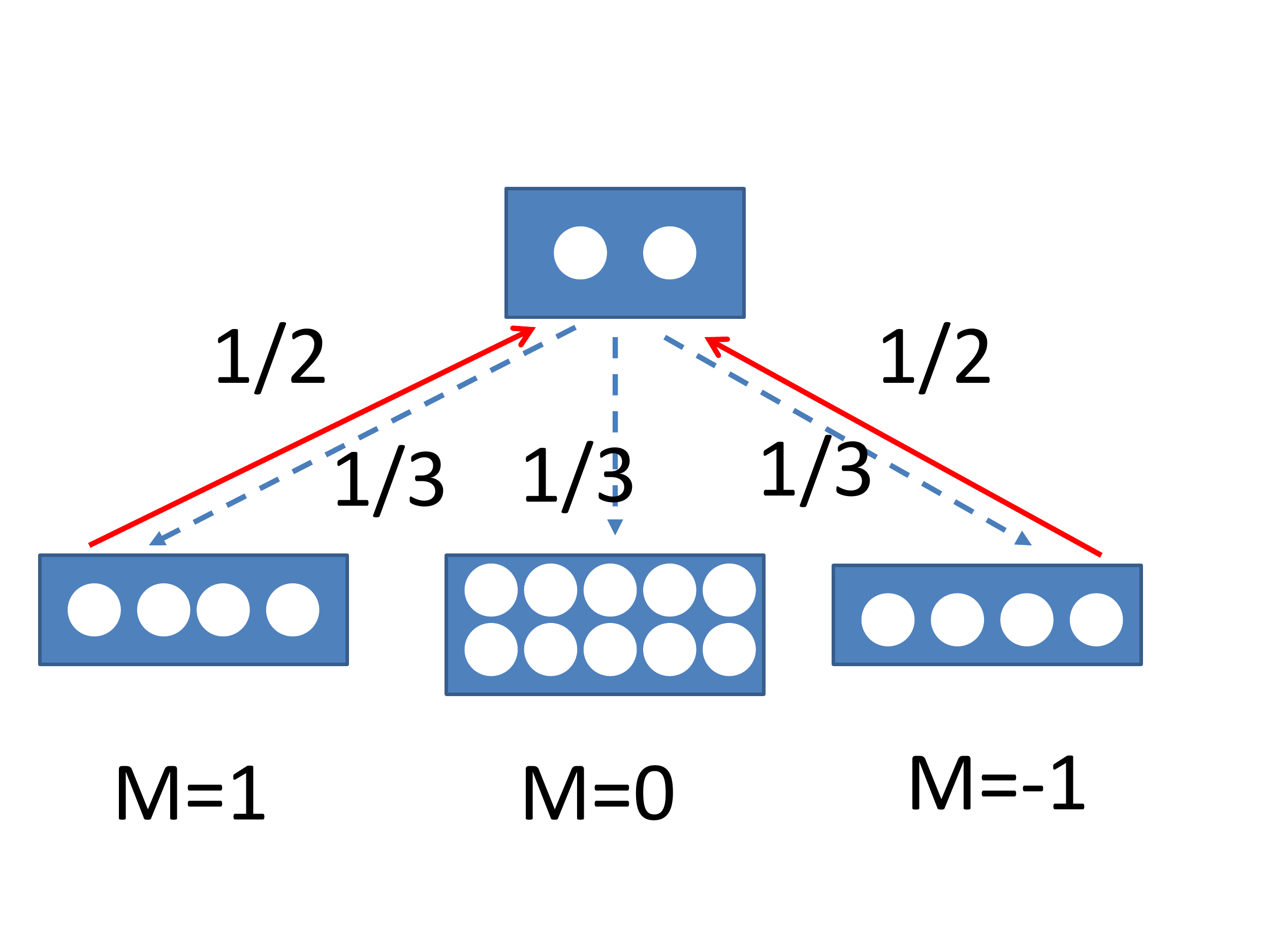}\label{fig1a}}
\subfigure[]{
 \includegraphics[width=0.41\columnwidth,
 height=0.22\textheight]{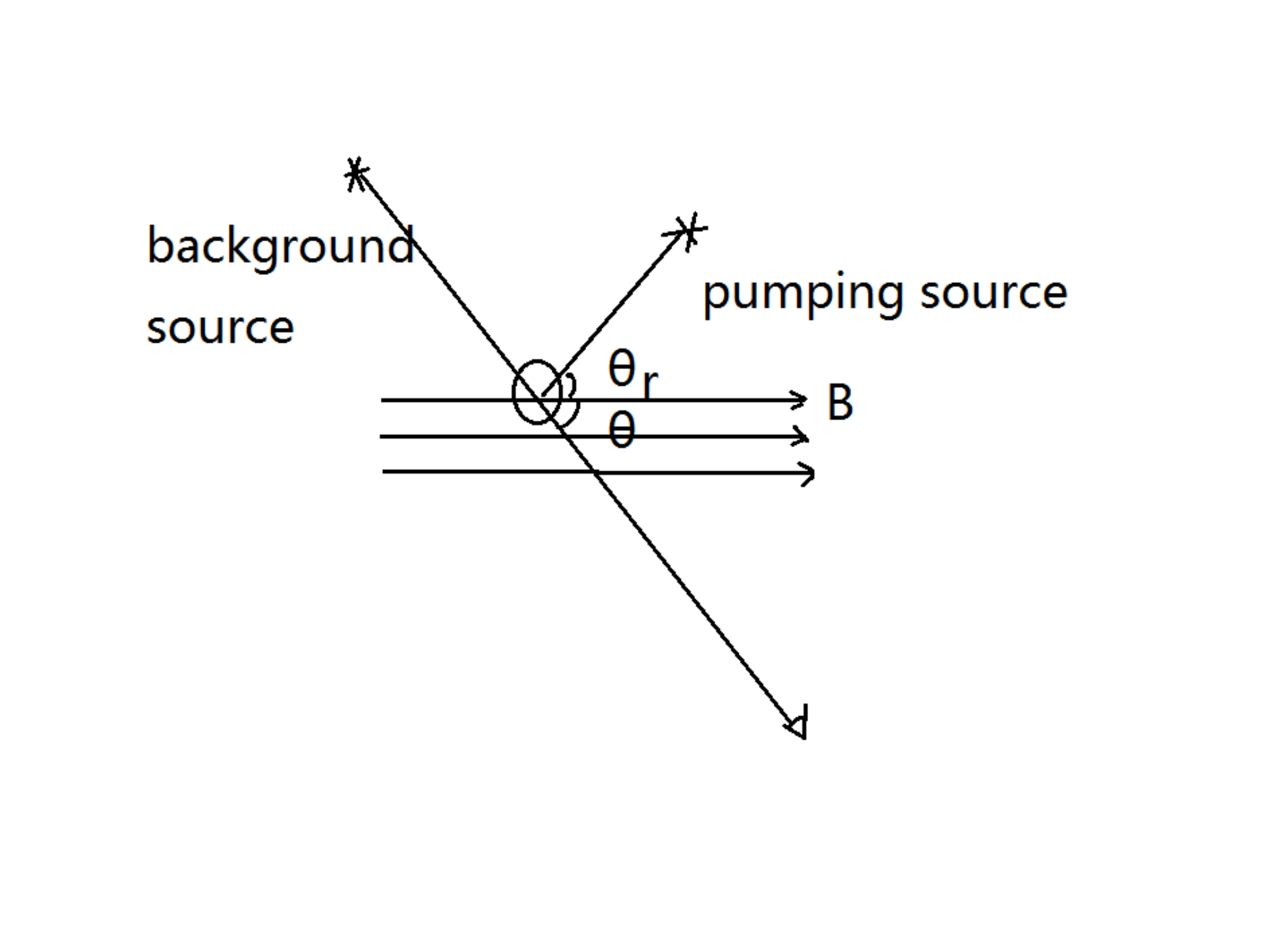}\label{fig1b}}
\caption{Diagram for the physics of GSA.
(a) Toy model for GSA. The atoms accumulate in the ground sublevel $M=0$ from $M=+1$ and $M=-1$ due to the pumping by anisotropic radiation.
(b) Typical environment where absorption line is polarized owing to GSA. Alignment is produced by a pumping source, which both influence the polarization of scattering(emission) lines and induce polarization on absorption of another background source from which the light passes through the aligned medium$^{\footnotesize 4}$. $\theta_r,\theta$ are the polar angles of the pumping radiation and line of sight measured in reference to magnetic field, respectively.}
\end{figure*}

An idealized toy model for GSA is presented in Fig.~\ref{fig1a} to demonstrate the basic physics \citep[see][]{YLfine}. It is important to note that {\em all the angles used in the paper are listed in Appendix~\ref{angles}.} As demonstrated in Fig.~\ref{fig1a}, a simplified atomic structure is exhibited with the total angular momentum $I=1$ on the ground state and $I=0$ on the upper state. $M$ is used to denote the projection of the angular momentum in the direction of the incident resonance photon beam. $M$ can be $-1, 0$, and $+1$ on the ground state, whereas $M$ can only be $0$ for the upper state. Photons in the unpolarized beam from the pumping source are equally left and right circularly polarized, and hence, induce transitions from the sublevels with $M =-1$ and $+1$ of the ground state to the upper state. On the other hand, atoms decay equally from the upper state to the sublevels with $M=-1, 0$, and $+1$ of the ground state. As a result, the atoms accumulate in the sublevel $M=0$ of the ground state from which no excitation is possible. Obviously, the alignment of the atoms in diffuse medium changes the optical properties (e.g. absorption) of the medium, leading to the change of the polarization observed, as demonstrated in Fig.~\ref{fig1b}.

The role of magnetic field is to mix different $M$ states, which is known as {\em magnetic realignment}. The atoms in the interstellar medium occupy differently on sublevels of the ground state influenced by magnetic field since the direction of the incoming photon generally does not coincide with magnetic field \citep[see][]{YLfine}. Clearly, the randomization in this situation is not complete and the magnetic realignment reflects the direction of magnetic field. It is important to note that magnetic realignment only happens when the magnetic precession rate is much higher than the excitation rate from the ground state. This is especially true for most diffuse medium, so GSA is an excellent magnetic tracer in diffuse medium.

In summary, for the atoms to be aligned, there must be enough degree of freedom (the quantum angular momentum number should be $\geqslant 1$) as well as the anisotropic incident radiation. It is important to note that in the study the collision excitation rate is lower than the radiative excitation rate, which is satisfied in many astrophysical environments such as the interplanetary medium, interstellar medium, the intergalactic medium, etc. The condition where GSA can be applied to trace magnetic field is defined as the magnetic realignment regime in \citet{2012JQSRT.113.1409Y}. The atomic alignment in diffuse medium in the existence of the extended radiation field can be different because the anisotropy of the extended radiation field is different from that of a point source.

\footnotetext[4]{There could be degenerate case where the pumping source is the same as the background source \citep[see][]{YLfine}.}

\subsection{Absorption and Emission}

It is proposed by \citet{YLfine} that the absorption line induced by GSA can be used to study astrophysical magnetic field. As illustrated in Fig.~\ref{fig1b}, the aligned atoms change the polarization of the photons from the background sources. For {\em absorption line} in the optically thin case, the Stokes parameters $S_i=[I,Q,U,V]$ are given by $Q=-\eta_{1}dI_{0}$, $I=I_{0}(1-\eta_{0}d)$, where the quantities $\eta_{i}$ are absorption coefficients defined in Eq.~\eqref{etai} in Appendix~\ref{parametersforeq}, the quantity $d$ is the thickness of the medium, and the quantity $I_{0}$ is the intensity of background. It is found that for unpolarized incoming light from the background sources, $U=V=0$, meaning that the polarization is linear, and can be only either parallel or perpendicular to magnetic field due to the fast magnetic precession. Thus, the degree of polarization $P$ per unit optical depth $\tau=\eta_{0}d$ is
\begin{equation}\label{ptau}
\frac{P}{\tau}=\frac{Q}{I\eta_{0}d} \approx \frac{1.5\sigma_{0}^{2}(J_l)\sin^{2}\theta \omega_{J_{l}J_{u}}^2}{\sqrt2+\sigma_{0}^{2}(J_l)(1-1.5\sin^{2}\theta)\omega_{J_{l}J_{u}}^2},
\end{equation}
where $\sigma_0^2$ depends on $\theta_r$ (the angle between the direction of the radiation field and magnetic field), $\theta$ is the angle between line of sight and magnetic field shown in Fig.~\ref{fig1b}, and the quantity $\omega_{J_lJ_u}^2$ is presented in Appendix~\ref{parametersforeq}. {\em $\sigma_0^2 \equiv \rho_0^2 / \rho_0^0$, the normalized dipole component of the ground state density matrix, is the actual measure of the ground state alignment.} The quantity $\rho_0^0$ is the total atomic population on the level, whereas the quantity $\rho^2_0$ is the dipole component of the density matrices\footnote[5]{For example, the irreducible tensor for $J/F=1$ is $\rho_0^2=[\rho(1,1)-2\rho(1,0)+\rho(1,-1)]$.}. Positive value of $P$ means the polarization being parallel to the direction of magnetic field, whereas a negative $P$ implies the case of perpendicular. Resulting from the sign reversal of $\sigma^2_0$, it is a general feature for the polarization $P$ in the magnetic realignment regime to flip between being parallel and perpendicular to magnetic field in unpolarized background sources. In reality, this means that the detection of any polarization in absorption line provides us with the magnetic field projected on the plane of sky with a $90^\circ$ degeneracy. This is all the information available from \citet{Goldreich:1982oq} effect which deals with the polarization of molecular radio lines arising from the magnetic redistribution on the upper levels instead of the ground state (see \citealt{YLfine,2012JQSRT.113.1409Y} for details). The $90^\circ$ degeneracy is also suggested in the grain alignment observed by Orion KL \citep[see][]{1998ApJ...502L..75R}. Additionally, if the degree of polarization is observed, both $\theta_r$ and $\theta$ can be determined, leading to removal of the $90^\circ$ degeneracy and the attainment of the 3D direction of magnetic field.

We only consider the influence of magnetic field on the ground state because the magnetic influence on the upper level is negligible when the magnetic precession period is comparable to the life time of the upper level. Emission line is influenced by GSA through the scattering from the aligned ground state. The linear polarization degree $P$ for optical thin case for emission line is given by
\begin{equation}
P=\sqrt{Q^2+U^2}/I=\sqrt{\epsilon_2^2+\epsilon_1^2}/\epsilon_0
\end{equation}
where $\epsilon_i$ are the emission coefficients given in Appendix~\ref{parametersforeq}.

In the paper, we mainly focus on the absorption line in extended radiation field. Only in \S 4 the polarization of emission line is discussed for circumstellar medium as an example since the effect of GSA is less straightforward for emission line to be compared with the observations. We emphasize that GSA can be applied to all the region discussed in this work to trace magnetic field with emission line.

\section{The formulae to compute GSA}

Atoms are excited by resonant radiation, resulting in photo-excitations and ensuing spontaneous emissions. The photo-excitation and magnetic precession determine the occupations among the sublevels of the ground state. The equations below describe the evolution of atom densities on both upper and ground states \citep[see][]{landi2004}:
\begin{equation}\label{upperevol}
\begin{split}
\dot{\rho}_q^k(J_u)&+2\pi i\nu_L g_u q\rho_q^k(J_u)=\\
&-\sum_{\substack{J_l}}A(J_u\rightarrow J_l)\rho_q^k(J_u)+\sum_{\substack{J_{l}k'}} [J_l]\left[\delta_{kk'}p_{k'}(J_u,J_l)B_{lu}\bar{J}^0_0+\sum_{\substack{Qq'}}r_{kk'}(J_u,J_l,Q,q')B_{lu}\bar{J}^2_Q\right]\rho^{k'}_{-q'}(J_l)
\end{split}
\end{equation}
\begin{equation}\label{groundevol}
\begin{split}
\dot{\rho}_q^k(J_l)+&2\pi i\nu_L g_l q\rho_q^k(J_l)=\\
&\sum_{\substack{J_u}}p_k(J_u,J_l)[J_u]A(J_u\rightarrow J_l)\rho_q^k(J_u)-\sum_{\substack{J_{u}k'}}\left[\delta_{kk'}B_{lu}\bar{J}^0_0+\sum_{\substack{Qq'}}s_{kk'}(J_u,J_l,Q,q')B_{lu}\bar{J}^2_Q\right]\rho^{k}_{q}(J_l)
\end{split}
\end{equation}
\begin{equation}
p_k(J_u,J_l)=(-1)^{J_u+J_l+1}\left\{\begin{array}{ccc}J_l&J_l&k\\J_u&J_u&1\end{array}\right\}, p_0(J_u,J_l)=\frac{1}{\sqrt{[J_u,J_l]}},
\end{equation}
\begin{equation}
r_{kk'}(J_u,J_l,Q,q)=(3[k,k',2])^{\frac{1}{2}}\left\{\begin{array}{ccc}1&J_u&J_l\\1&J_u&J_l\\2&k&k'\end{array}\right\}\left(\begin{array}{ccc}k&k'&K\\q&q'&Q\end{array}\right),
\end{equation}
\begin{equation}
s_{kk'}(J_u,J_l,Q,q)=(-1)^{J_l-J_u+1}[J_l](3[k,k',K])^\frac{1}{2}\left(\begin{array}{ccc}k&k'&2\\q&q'&Q\end{array}\right) \left\{\begin{array}{ccc}1&1&2\\J_l&J_l&J_u\end{array}\right\}\left\{\begin{array}{ccc}k&k'&2\\J_l&J_l&J_l\end{array}\right\}.
\end{equation}
where the quantities $J_{u}$ and $J_{l}$ are the total angular momentum quantum numbers for the upper levels and lower levels, respectively. The quantity $A$ is the Einstein spontaneous emission rate and the quantity $B$ is the Einstein coefficient for absorption and stimulated emission, as defined in Appendix~\ref{parametersforeq}\footnote[6]{The data of Einstein coefficients used in the paper are taken from the Atomic Line List (\url{http://www.pa.uky.edu/~peter/atomic/}) and the NIST Atomic Spectra Database.}. The quantities $\rho_q^k$ and $\bar{J}_Q^K$, both defined in Appendix~\ref{densitymat}, are irreducible density matrices for the atoms and the radiation field, respectively. $6-j$ and $9-j$ symbols are represented by the matrices with $"\{$ $\}"$, whereas $3-j$ symbols are indicated by the matrices with $"()"$ (see \citealt{1989PhT....42l..68Z} for details). The symbol $[j]$ represents the quantity $2j+1$, e.g., $[J_l]=2J_l+1$, $[J_l,J_u]=(2J_l+1)(2J_u+1)$, etc. The evolution of the upper state $[\rho_q^k(J_u)]$ is illustrated in Eq.~\eqref{upperevol}, whereas Eq.~\eqref{groundevol} describes the ground state $[\rho_q^k(J_l)]$. The second terms on the left side of Eq.~\eqref{upperevol} and Eq.~\eqref{groundevol} represent the magnetic realignment, which can be neglected on the upper levels. The two terms on the right side represent spontaneous emissions and the excitations from the ground level. Transitions to all upper states and to all ground sublevels are counted by summing up $J_u$ and $J_l$. Note that the quantities $k$ and $q$ are conserved for the symmetric processes of spontaneous emission and magnetic realignment.

The excitation depends on
\begin{equation}\label{tensordef}
\bar{J}_Q^K=\int d\nu\frac{\nu_0^2}{\nu^2}\xi(\nu-\nu_0)\oint\frac{d\Omega}{4\pi}\sum_{\substack{i=0}}^{\substack{3}} \bar{J}_Q^K(i,\Omega)S_i(\nu,\Omega),
\end{equation}
where $\bar{J}_Q^K$ is the radiation tensor of the incoming light averaged over the whole solid angle and line profile $\xi(\nu-\nu_0)$. The irreducible unit tensors for the Stokes Parameters $I$, $Q$, and $U$ are:
\begin{equation}\label{orintensor}
\begin{split}
\mathcal{J}_0^0(i,\Omega)=\left(
\begin{array}{c}
1\\
0\\
0\\
\end{array}\right),
&\mathcal{J}_0^2(i,\Omega)=\frac{1}{\sqrt{2}}\left[
\begin{array}{c}
1-1.5\sin^{2}\theta\\
-3/2\sin^{2}\theta\\
0\\
\end{array}\right],\\
\mathcal{J}_{\pm2}^{2}(i,\Omega)=\sqrt{3}e^{\pm 2i\phi}\left[
\begin{array}{c}
\sin^{2}\theta/4\\
-(1+\cos^{2}\theta)/4\\
\mp i\cos\theta/2\\
\end{array}\right],
&\mathcal{J}_{\pm1}^{2}(i,\Omega)=\sqrt{3}e^{\pm i\phi}\left(
\begin{array}{c}
\mp \sin 2\theta/4\\
\mp \sin 2\theta/4\\
-i \sin \theta/2\\
\end{array}\right).
\end{split}
\end{equation}

The nonzero elements of the radiation tensors for the incoming radiation $(\theta_r,$ $\phi_r)$ are obtained by substituting Eq.~\eqref{orintensor} into Eq.~\eqref{tensordef}:
\begin{equation}
\begin{split}
\bar{J}_0^0=I_*,
&\bar{J}_0^2=\frac{W_a}{2\sqrt{2}W}(2-3\sin^{2}\theta)I_*,\\
\bar{J}_{\pm2}^{2}=\frac{\sqrt{3}W_a}{4W}\sin^{2}\theta I_* e^{\pm 2i\phi},
&\bar{J}_{\pm1}^{2}=\mp\frac{\sqrt{3}W_a}{4W}\sin 2\theta I_* e^{\pm i\phi},
\end{split}
\end{equation}
where the quantity $W$, composed of an anisotropic part $W_a$ and an isotropic part $W_i$, is the dilution factor of the radiation field. Clearly, $W_i$ is nonzero for extended radiation field, while it is zero for a beam of light. Therefore, both the alignment and the polarization in extended radiation field are expected to be different compared to the case of a point source. $I_*$ is the intensity of the radiation source averaged over the whole solid angle. For example, the intensity of the blackbody radiation source is given by:
\begin{equation}\label{bbradiation}
I_*=W\frac{2h\nu^3}{c^2}\frac{1}{e^\frac{h\nu}{k_BT}-1}
\end{equation}

Hence, the steady state occupations of atoms on the ground state is obtained by setting the left side of Eq.~\eqref{upperevol} and Eq.~\eqref{groundevol} zero:
\begin{equation}\label{groundoccu}
\begin{split}
2\pi i\rho_q^k(J_l)qg_l\nu_L&-\sum_{\substack{J_uk'}}\bigg\{p_k(J_u,J_l)\frac{[J_u]}{\sum_{J''_{l}}A''/A+i\Gamma'q} \sum_{\substack{J'_l}}B_{lu}[J'_l]\left[\delta_{kk'}p_{k'}(J_u,J'_l)\bar{J}^0_0+\sum_{\substack{Qq'}}r_{kk'}(J_u,J_l,Q,q')\bar{J}^2_Q\right]\rho^{k'}_{-q'}(J_l)\\
&-\left[\delta_{kk'}B_{lu}\bar{J}^0_0+\sum_{\substack{k'Qq'}}B_{lu}s_{kk'}(J_u,J_l,Q,q')\bar{J}^2_Q\right]\rho^{k}_{-q}(J_l)\bigg\}=0
\end{split}
\end{equation}
where $\Gamma'$ equals $2\pi\nu_L g_l/A$.

Owning to the fast magnetic procession, all Zeeman coherence components $(q \neq 0)$ disappear \citep[see][]{YLfine}. Moreover, only $\bar{J}_{0}^{0,2}$ components are involved due to the selection rule of $3-j$ symbols. Then it can be deduced from Eq.~\eqref{groundoccu} that $\sigma_0^2\equiv \rho^2_0/\rho^0_0 \propto \bar{J}_0^2$. Thus the degree of polarization $P$ is proportional to the radiation tensor $\bar{J}_0^2$ (see Eq.~\ref{ptau}).

The photons are from different directions in extended radiation field. Therefore, it is necessary to integrate the radiation tensor over the whole solid angle distribution of the extended radiation field\footnote[7]{In earlier studies, the distribution of the radiation source with the point source assumption is a $\delta-function$, in accordance.} (see Eq.~\ref{tensordef}). Specifically, the extended radiation source can be treated as an aggregation of point sources if the radiation source is identified. Alternatively, the radiation field can be decomposed to study the alignment parameter in each multipole component in the case where individual radiation source cannot be pinpointed, e.g., in the interstellar radiation field.

\section{GSA in circumstellar medium}

\begin{figure*}
\centering
 \subfigure[]{
\includegraphics[width=0.32\columnwidth,
 height=0.22\textheight]{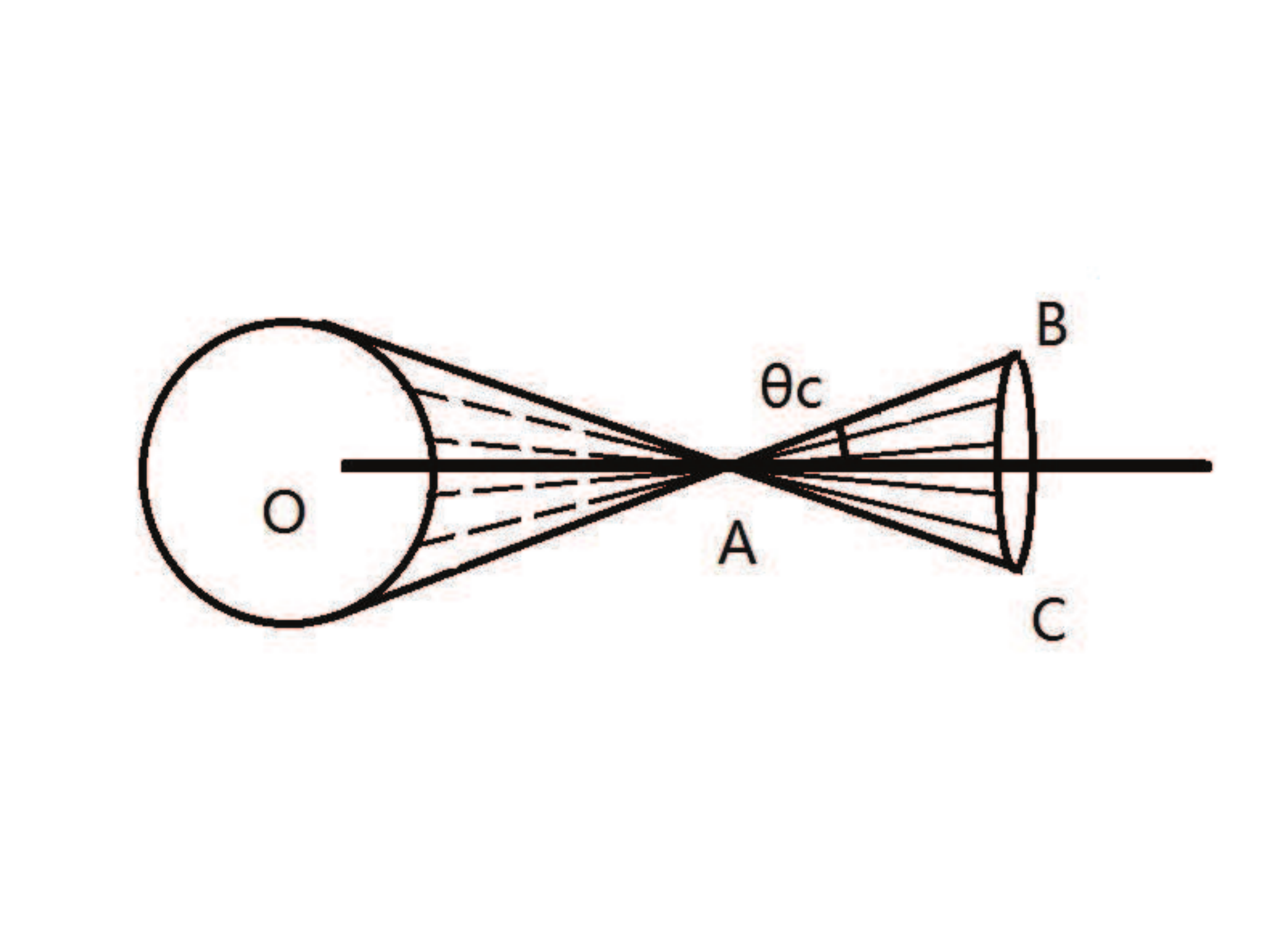}\label{fig2a}}
\subfigure[]{
\includegraphics[width=0.32\columnwidth,
 height=0.22\textheight]{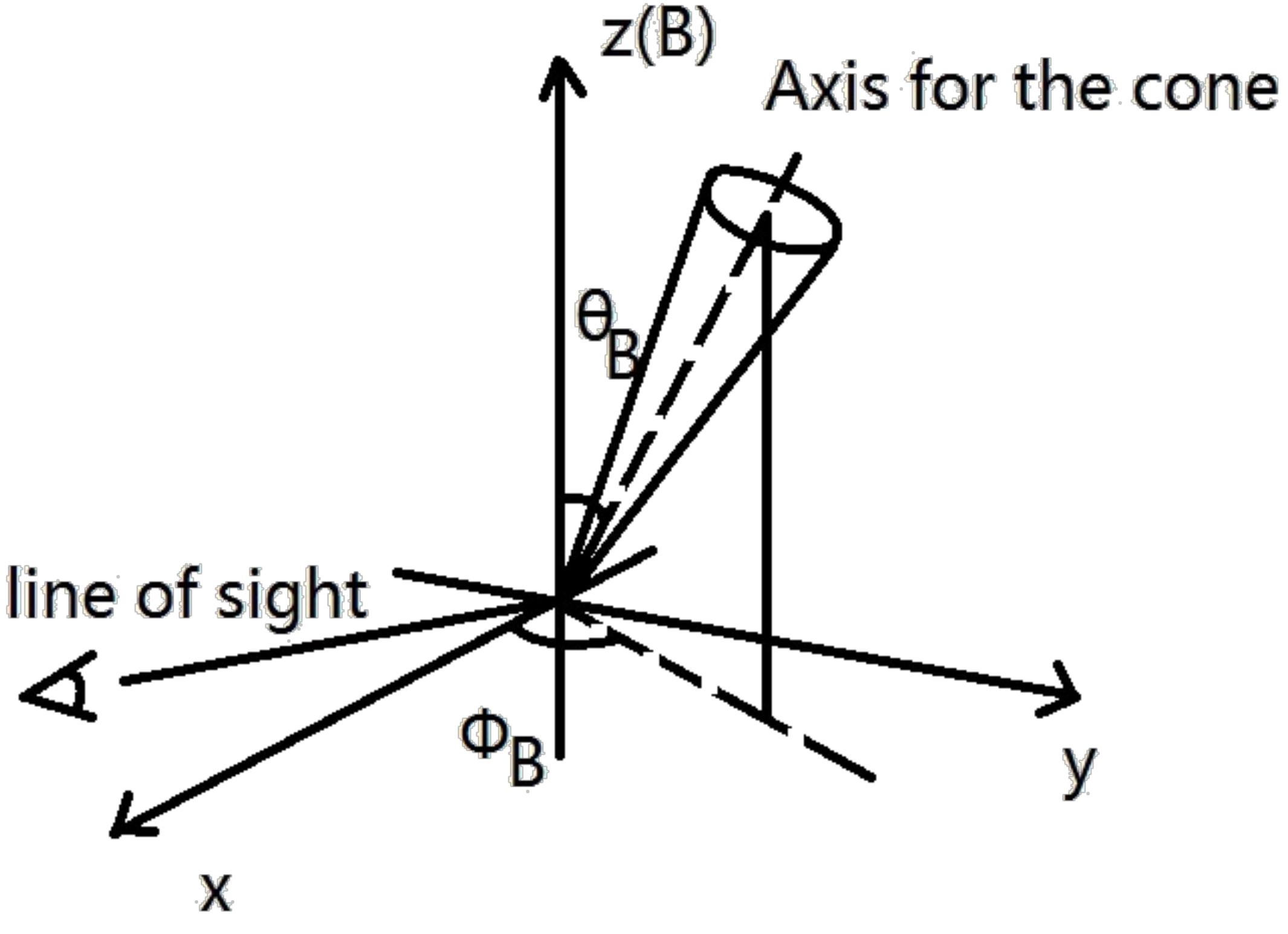}\label{fig2b}}
\subfigure[]{
\includegraphics[width=0.32\columnwidth,
 height=0.22\textheight]{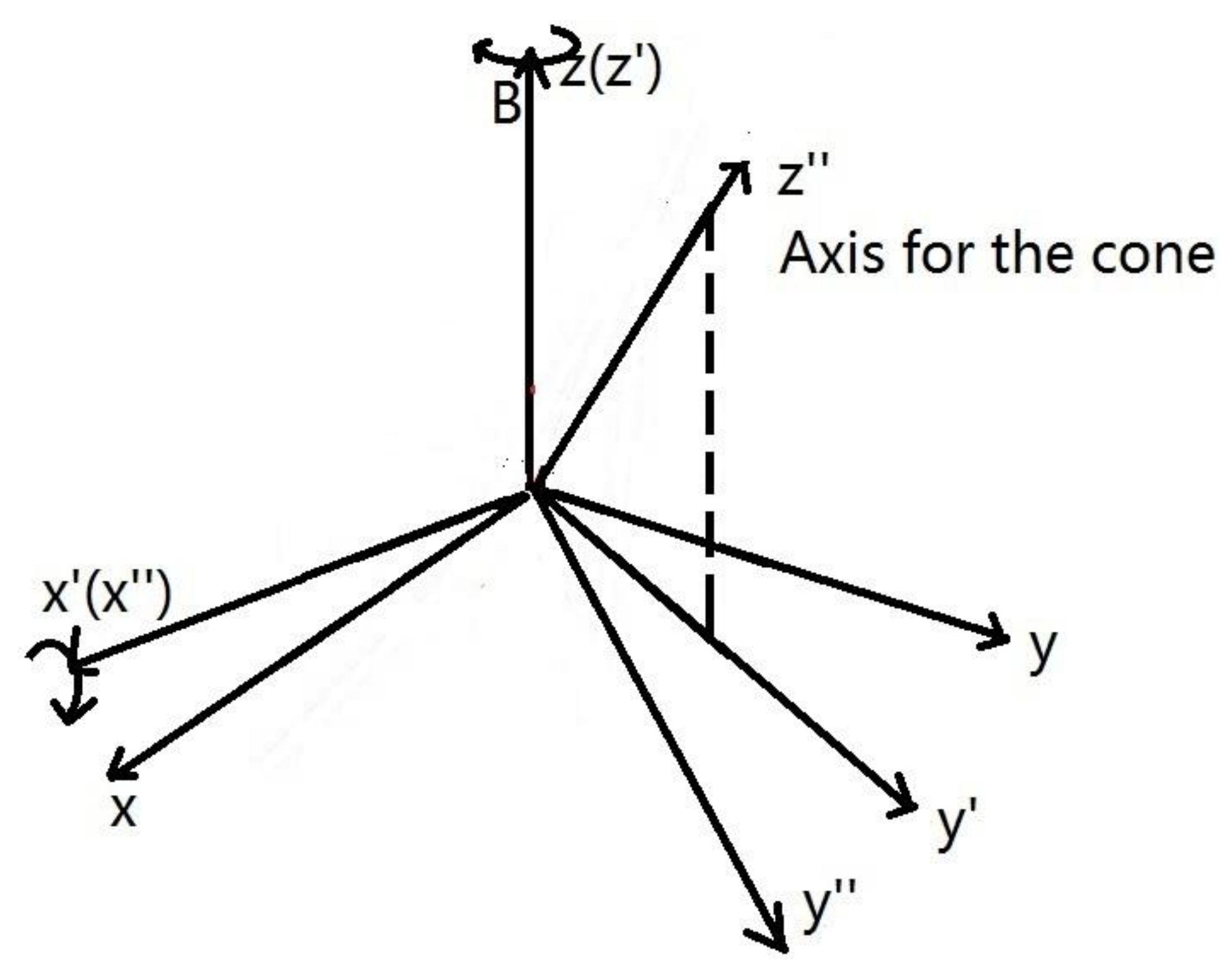}\label{fig2c}}
\caption{Geometry of the radiation field in circumstellar medium
(a) Radiation field as a light cone A-BC. A: the diffuse medium where the alignment happens; Sphere O: the extended radiation source; $\theta_c$: the cone angle.
(b) Original coordinate frame. $z$ axis: the direction of magnetic field; line of sight is in the $x-z$ plane; ($\theta_B,$ $\phi_B$): the angle coordinate for the axis of the cone in this frame.
(c) Coordinate conversion. $xyz-$frame: original coordinate frame defined in Fig.~\ref{fig2b}; $x'y'z'-$frame: the original frame rotating around the $z-$axis so that the axis of the cone is in the $x'y'-$plane; $x''y''z''-$frame: $x'y'z'-$frame rotating around the $x'-$axis so that $z''-$axis is the axis of the cone.}
\end{figure*}

A geometric model is proposed in Fig. 2 to illustrate GSA in the circumstellar region where a star, the dominant radiation source, is not distant enough to be treated as a point source. The incoming radiation from the star O to the medium region A forms a cone A-BC with the symmetric axis of line OA, as demonstrated in Fig.~\ref{fig2a}. The cone angle $\theta_c$ is dependent on the radius of the star and the distance between the star center and the medium A. The radiation frame $(xyz)$ and angle coordinate ($\theta_B,$ $\phi_B$) are defined in Fig.~\ref{fig2b}. However, for the sake of simplicity, the calculation is performed in the coordinate system $(x''y''z'')$, as shown in Fig.~\ref{fig2c}. The mathematics involved in the coordinate conversion is demonstrated in Appendix~\ref{coordinatestransfer}.

\begin{figure*}
\centering
\subfigure[]{
\includegraphics[width=0.45\columnwidth,
 height=0.28\textheight]{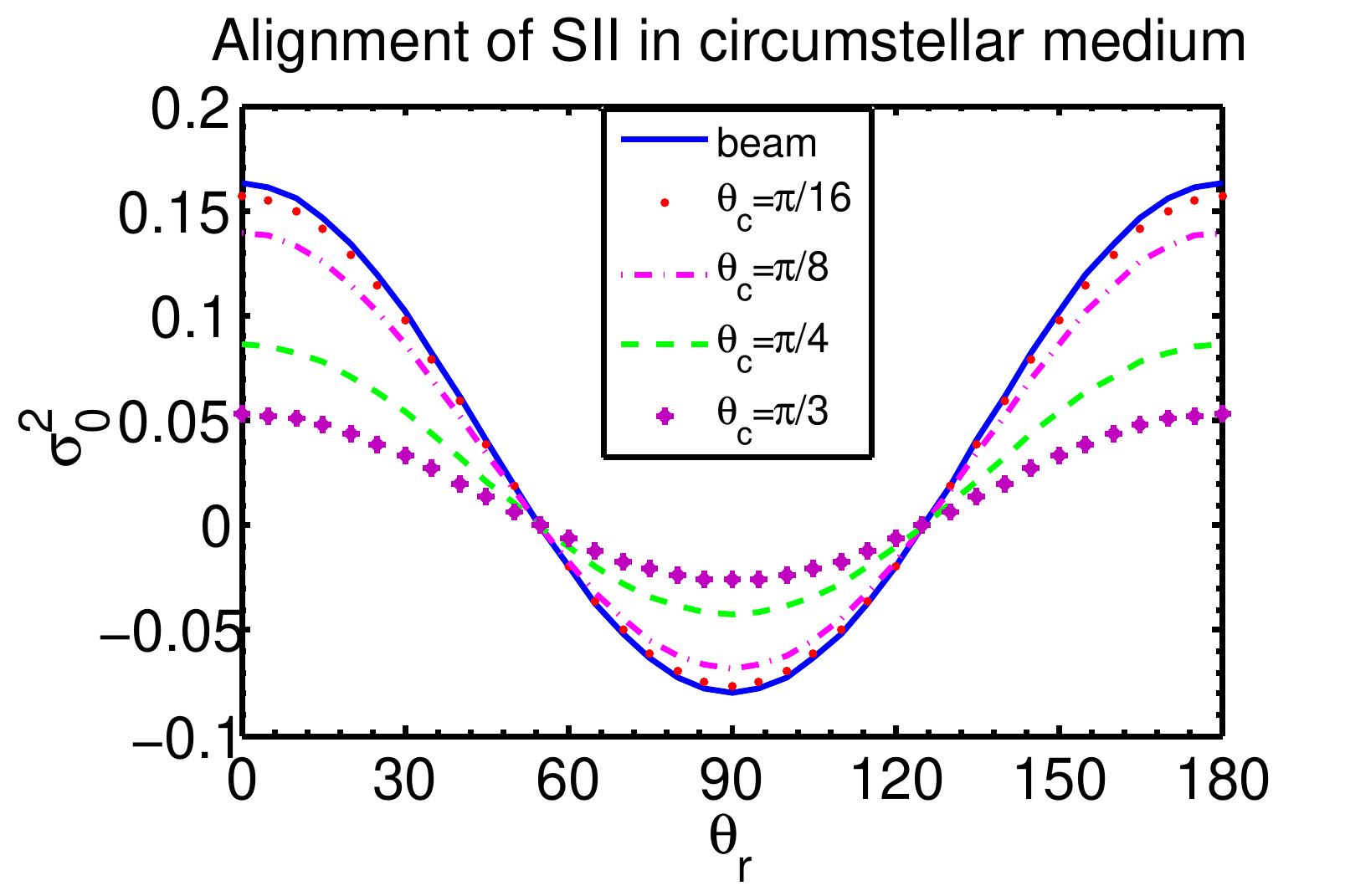}\label{fig3a}}
\subfigure[]{
\includegraphics[width=0.45\columnwidth,
 height=0.28\textheight]{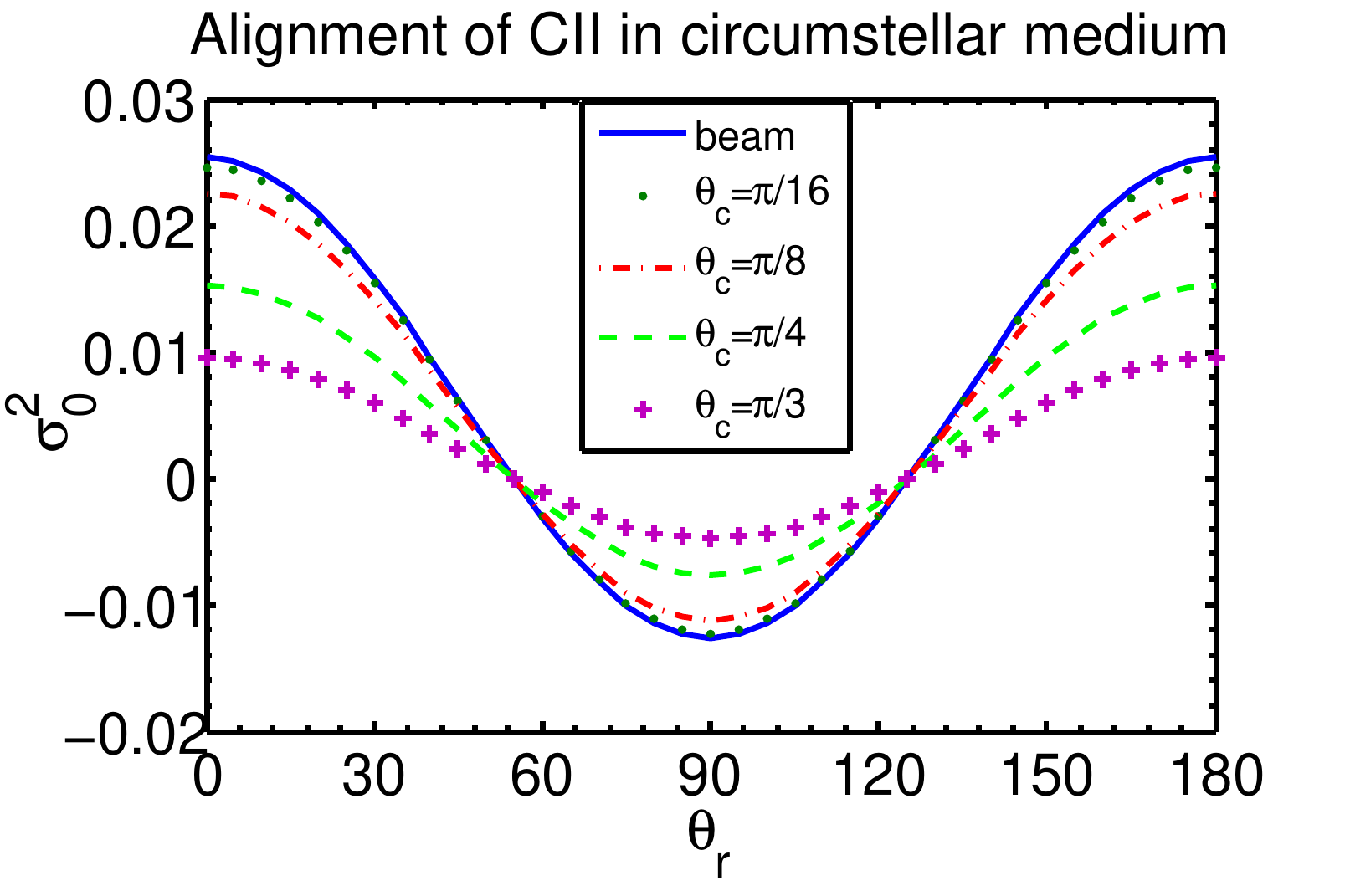}\label{fig3b}}
\caption{GSA in the vicinity of one dominant source.
$\theta_r$: the angle between magnetic field and the direction of the radiation cone axis; $\theta_c$: the size of the radiation cone defined in Fig.~\ref{fig2a}. Five lines represent a beam, light cone with $\theta_c=\pi/16$, $\theta_c=\pi/8$, $\theta_c=\pi/4$, and $\theta_c=\pi/3$, respectively.
(a) Alignment of SII; (b) Alignment of CII.}
\end{figure*}

The alignment in this region is compared with the case of the point radiation source in Fig. 3. For absorption line, we use the elements of SII and CII as examples in the paper without the loss of generality. The case of SII absorption line in a beam of light has been demonstrated in \citet{YLfine}. The ground state of SII is $4S_{\frac{3}{2}}^0$ ($J_l=\frac{3}{2}$) and the upper states are $4P_{\frac{1}{2}; \frac{3}{2}; \frac{5}{2}}$ ($J_u=\frac{1}{2}, \frac{3}{2}, \frac{5}{2}$). Since we consider the more general situation where the incoming light is unpolarized, the nonzero density matrices of the ground state are those with $k=0,2$.

Using the formulae presented in \S 3, we obtain the alignment parameter of SII in circumstellar medium:
\begin{equation}\label{spheredens}
\sigma_{0}^{2}(J_l)=\frac{-1.7141f_{c}(\cos\theta_B,\cos\theta_c)}{21.3806+0.2027f_{c}(\cos\theta_B,\cos\theta_c)},
\end{equation}
in which
\begin{equation}
f_{c}(x,y)=\frac{1}{2}(3x^{2}-1)(-y^{3}+y).
\end{equation}

By substituting Eq.~\eqref{spheredens} into Eq.~\eqref{ptau}, the polarization for SII absorption line in circumstellar medium is obtained:
\begin{equation}
\frac{P}{\tau}=\frac{-2.5711f_{c}(\cos\theta_B,\cos\theta_c)\sin^{2}\theta \omega_{J_lJ_u}^2}{30.2367+0.2867f_{c}(\cos\theta_B,\cos\theta_c)-1.7141f_{c}(\cos\theta_B,\cos\theta_c)(1-1.5\sin^{2}\theta)\omega_{J_lJ_u}^2}.
\end{equation}
The results of SII in circumstellar medium with different cone angles are shown in Fig.~\ref{fig3a}.

Atom species such as CII have more than one ground sublevels, which are different from SII. The absorptions from the two sublevels of the ground state $2P_{1/2,3/2}^{0}$ to the upper states $2S_{1/2}$, $2D_{3/2,5/2}$ are considered. Only the sublevel $P_{3/2}^{0}$ on the ground state is alignable with two density tensors $\rho_{0}^{0,2}$. Results of CII in circumstellar medium with different cone angles are plotted in Fig.~\ref{fig3b}.

Fig. 3 indicates that the polarization of absorption line induced by GSA flips between being parallel and perpendicular to magnetic field when $\theta_B$ is the Van Vleck angle $54.7^{\circ}$ \citep{1925PNAS...11..612V,1974PASP...86..490H}, resulting from the coupling of two oscillators in the $x-y$ plane caused by Larmor precession around magnetic field. This switch is a general feature regardless of the specific atomic species if the background source is unpolarized. Furthermore, the alignment in circumstellar medium approaches to that in a beam of light \citep[see][]{YLfine} as $\theta_{c}\rightarrow 0$, as demonstrated in Fig. 3. Apparently, the radiation source can be considered as a point source when $\theta_c$ approaches to zero, i.e., when the radiation source is sufficiently distant.

\begin{figure*}
\centering
 \subfigure[]{
\includegraphics[width=0.45\columnwidth,
 height=0.28\textheight]{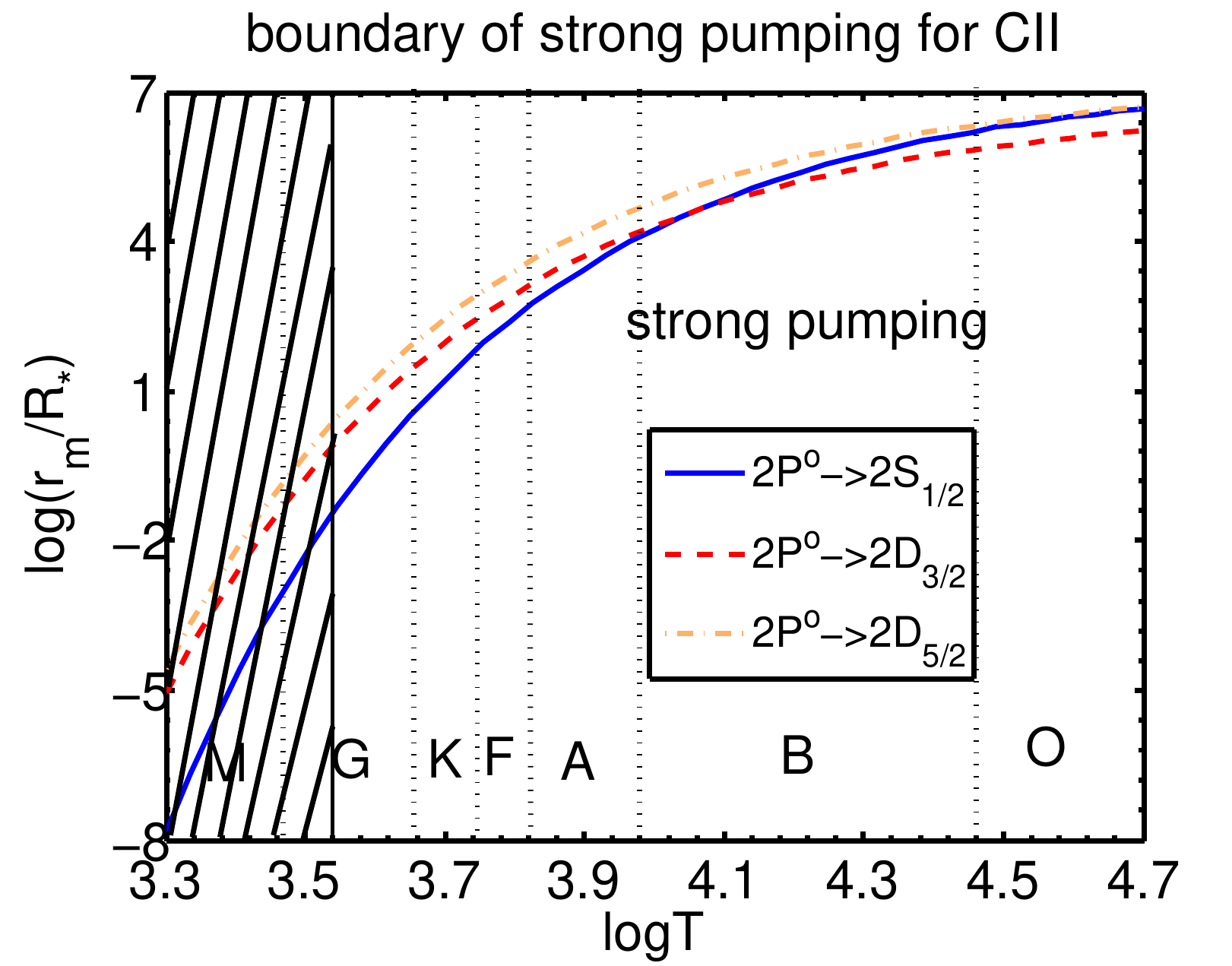}\label{fig4a}}
\subfigure[]{
\includegraphics[width=0.45\columnwidth,
 height=0.28\textheight]{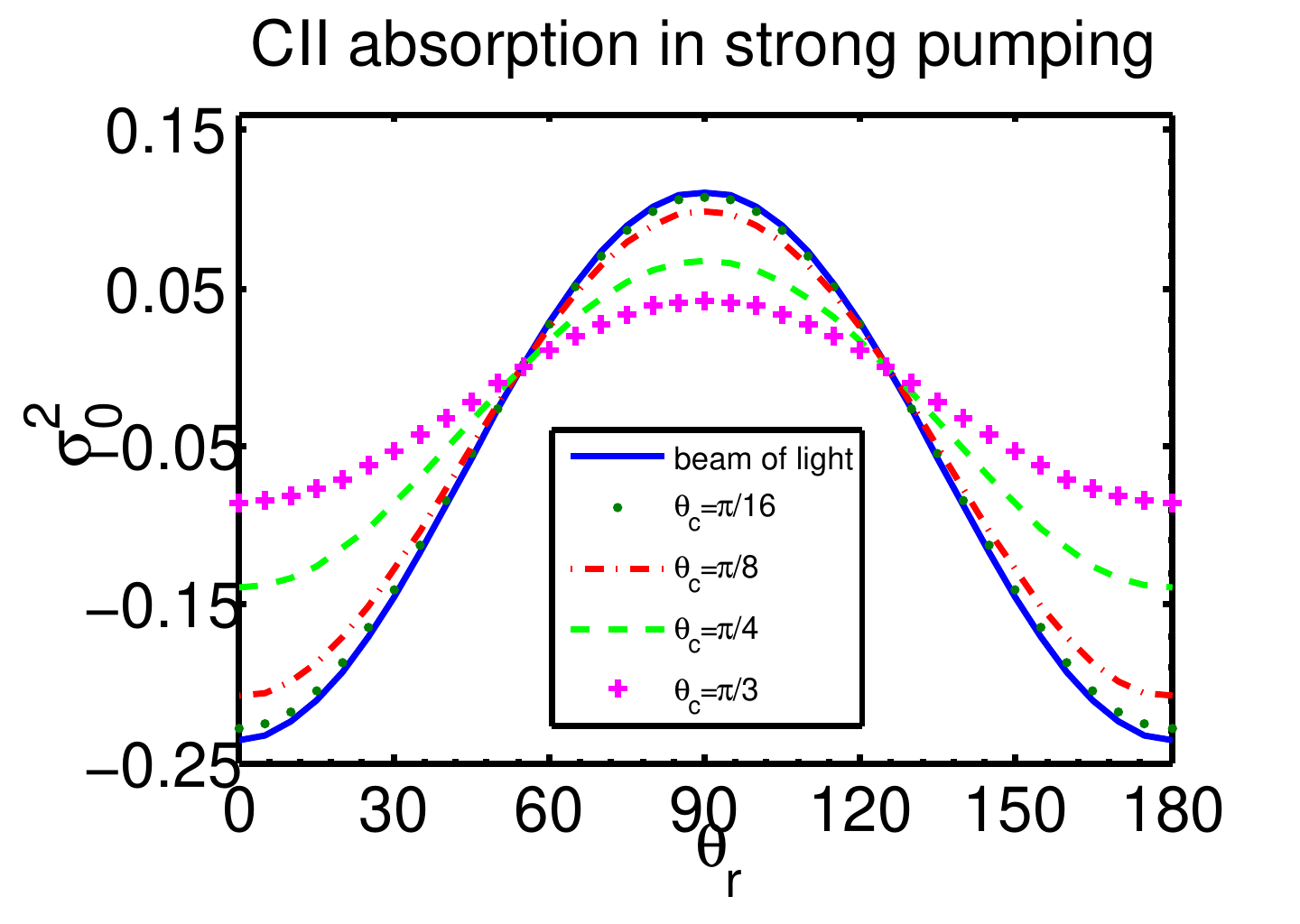}\label{fig4b}}
\caption{
(a) The strong pumping regime for the alignment of CII induced by different type of stars. $r_m$: the distance from the diffuse medium to the radiation source; $R_*$: the radius of the pumping star; $x-$axis: the temperature of the radiation source; Different lines illustrate the boundaries for different transitions. Apparently, $r_m$ is meaningful only if $log(\frac{r_m}{R_*})>0$. Therefore, the strong pumping approximation cannot be applied to the shaded area. The region where the strong pumping approximation can be applied is noted. As an example, the alignment in the vicinity of B type stars with the temperature $T\sim15000K$ is presented in (b) for strong pumping approximation. Five lines represent a beam, light cone with $\theta_c=\pi/16$, $\theta_c=\pi/8$, $\theta_c=\pi/4$, and $\theta_c=\pi/3$, respectively.}
\end{figure*}

It has been illustrated in \citet{YLfine} that the magnetic dipole transition has to be taken into account for the alignment of atoms with multiple ground levels in the weak pumping regime, where the medium is distant enough from the radiation source, i.e., the magnetic dipole radiation rate $A_m$ is comparable to the pumping rate $\tau^{-1}_R$. Conversely, the magnetic dipole can be neglected in strong pumping regime. The boundary of the two regime $r_m$ is defined as the place where
\begin{equation}\label{radiusdef}
A_m\sim\tau_R^{-1}=\left(\frac{R_*}{r_m}\right)^2BI_*,
\end{equation}
in which $R_*$ is the radius of the star. $A_m$ is the Einstein coefficient of the magnetic dipole transition on the ground state. Inserting Eq.~\eqref{bbradiation} into Eq.~\eqref{radiusdef}, we obtain the radius of the boundary sphere inside which is the strong pumping regime:
\begin{equation}\label{radiuscal}
\frac{r_m}{R_*}=\sqrt{W\frac{[J_u]}{[J_l]}\frac{A(J_u\rightarrow J_l)}{A_m}\frac{1}{e^\frac{h\nu}{k_BT}-1}}.
\end{equation}
The boundary for CII is illustrated in Fig.~\ref{fig4a}. Obviously, the boundary varies with the type of the central star. The strong pumping approximation is valid for the regime noted in Fig.~\ref{fig4a}. An example is provided in Fig.~\ref{fig4b} for B type star with the temperature of $15000K$, revealing the alignment of CII with different cone angle $\theta_c$ in the strong pumping approximation.

\begin{figure*}
\centering
\subfigure[]{
 \includegraphics[width=0.45\columnwidth,
 height=0.28\textheight]{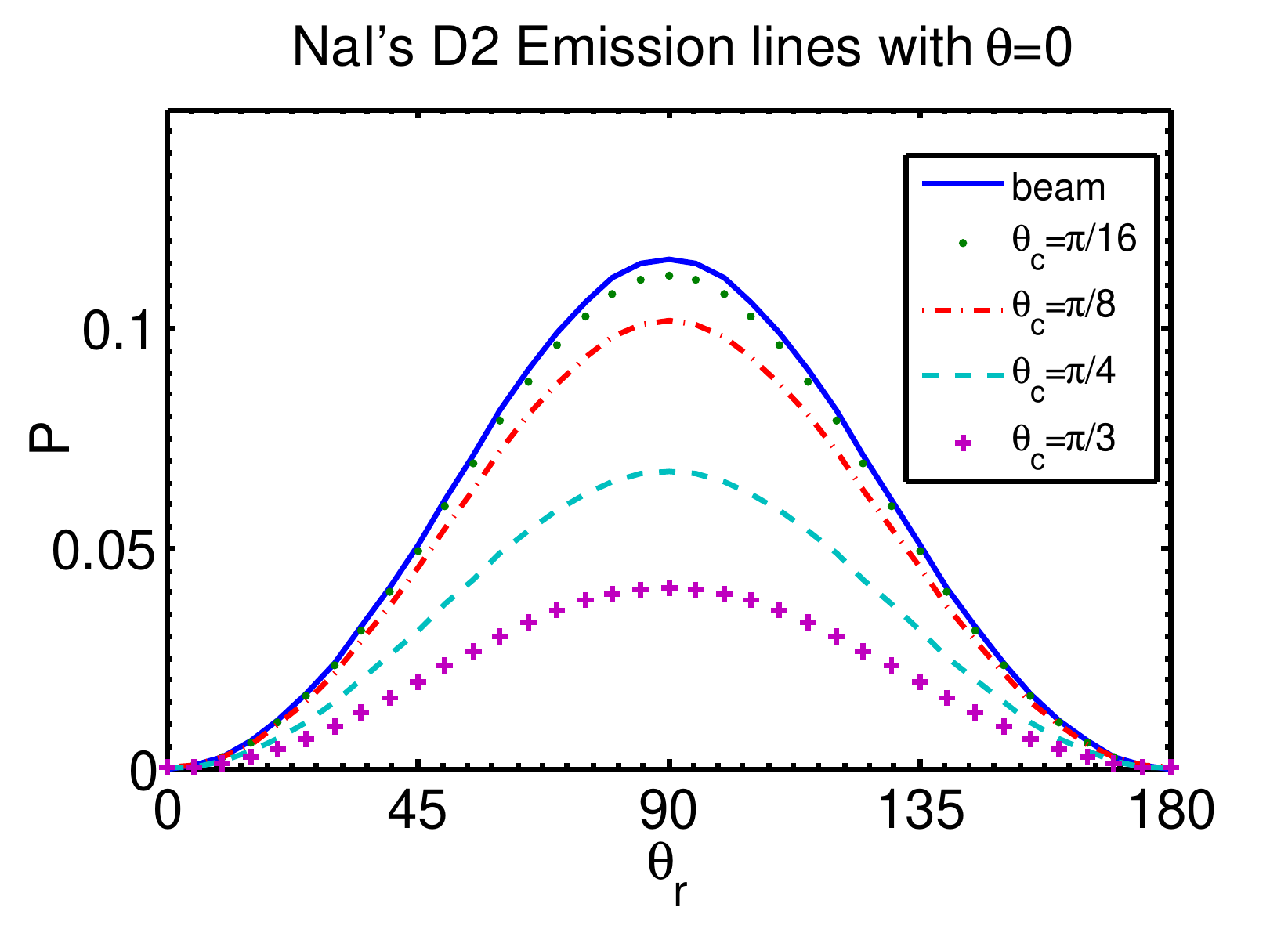}\label{fig5a}}
 \subfigure[]{
 \includegraphics[width=0.45\columnwidth,
 height=0.28\textheight]{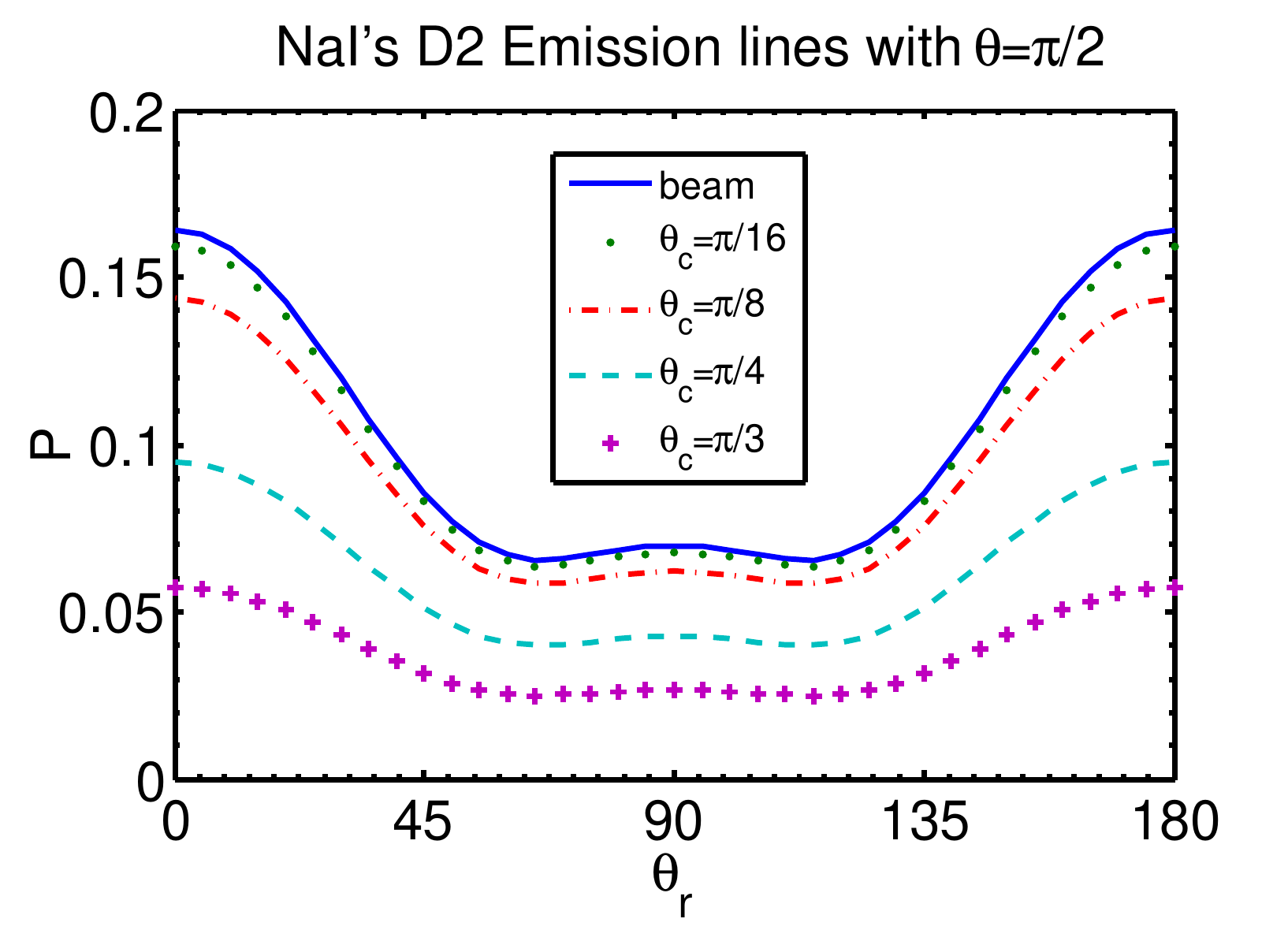}\label{fig5b}}
\caption{Polarization of NaI's D2 emission line in the vicinity of one dominant source influenced by GSA. Five lines represent the cases of a beam, light cone with $\theta_c=\pi/16$, $\theta_c=\pi/8$, $\theta_c=\pi/4$, and $\theta_c=\pi/3$ for (a) $\theta=0$ and (b) $\theta=\pi/2$, respectively.}
\end{figure*}

The polarization of $D2$ emission line of NaI in circumstellar medium with different cone angle $\theta_c$ are compared with that in a beam of light, as shown in Fig.~\ref{fig5a} and Fig.~\ref{fig5b} for $\theta=0$ and $\theta=\pi/2$, respectively. Obviously, the linear polarization of NaI's $D2$ emission line approaches to the case with a point source \citep[see][]{YLhyf} as $\theta_{c}\rightarrow0$.

It is demonstrated in this section that the radiation source of circumstellar medium can be treated as a point source when it is far away from the medium. In addition, the switch point of the polarization between being parallel and perpendicular to magnetic field is the Van Vleck angle in circumstellar medium, similar to that in a point source. {\em Therefore, 2D magnetic field in circumstellar medium can be directly determined by the direction of polarization with a $90^{\circ}$ degeneracy; if the degree of polarization are observed, the 3D magnetic field can be identified.}


\section{GSA in binary systems}

In this section we consider GSA in binary systems, which are quite common in the universe. Studies since the early 19th century have suggested that a large fraction of stars are in binary system. The binary systems are now hotly discussed in studies of star evolution (e.g., see \citealt{Sirius}). Moreover, binaries can be surrounded by circumbinary disks, which may be important for star formation \citep{2000prpl.conf..703M}. Examples for circumbinary disks can be found in \citet{1991ApJ...370L..35A}.

\begin{figure*}
\centering
\subfigure[]{
\includegraphics[width=0.47\columnwidth,
 height=0.26\textheight]{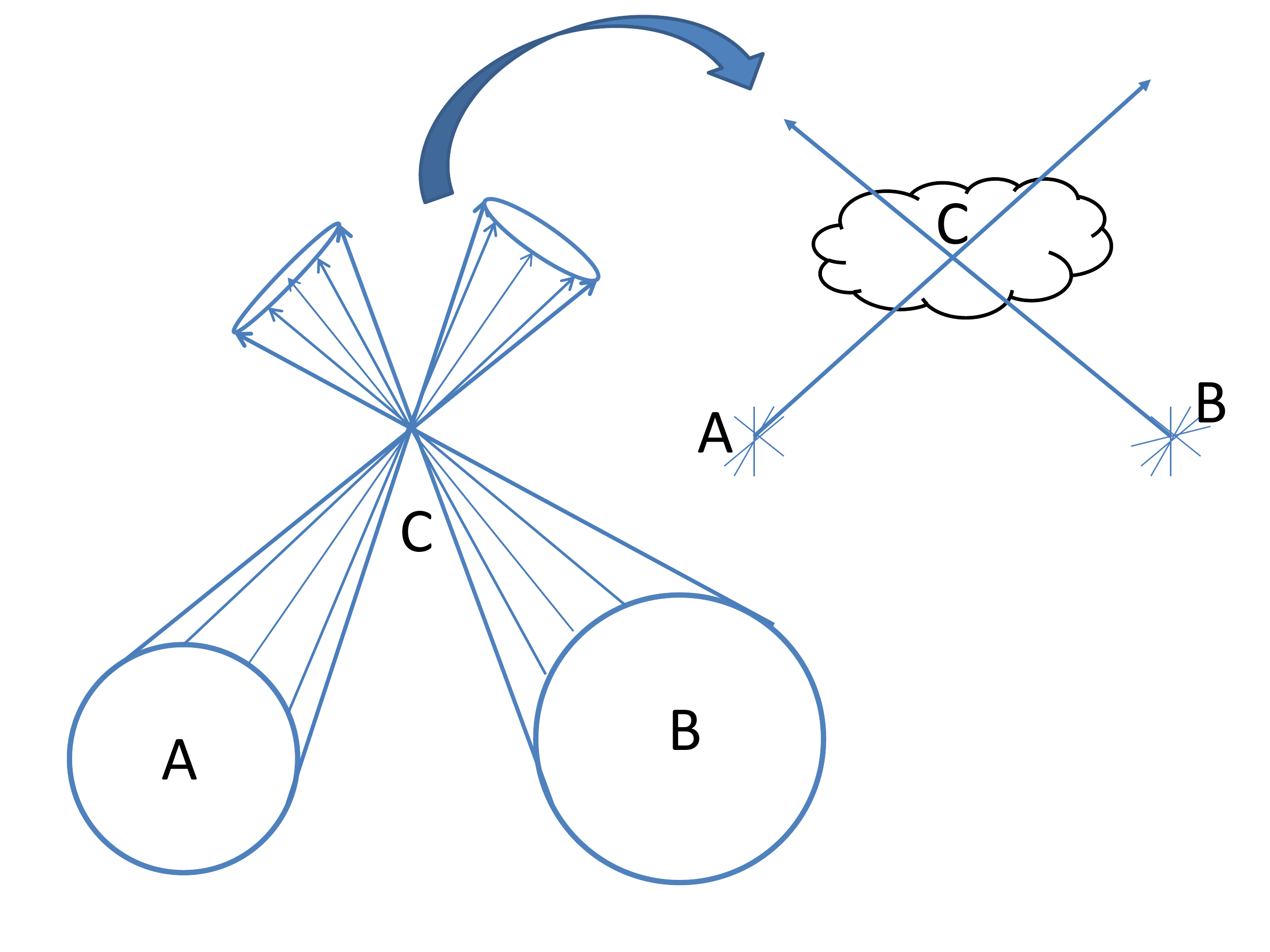}\label{fig6a}}
\subfigure[]{
\includegraphics[width=0.47\columnwidth,
 height=0.26\textheight]{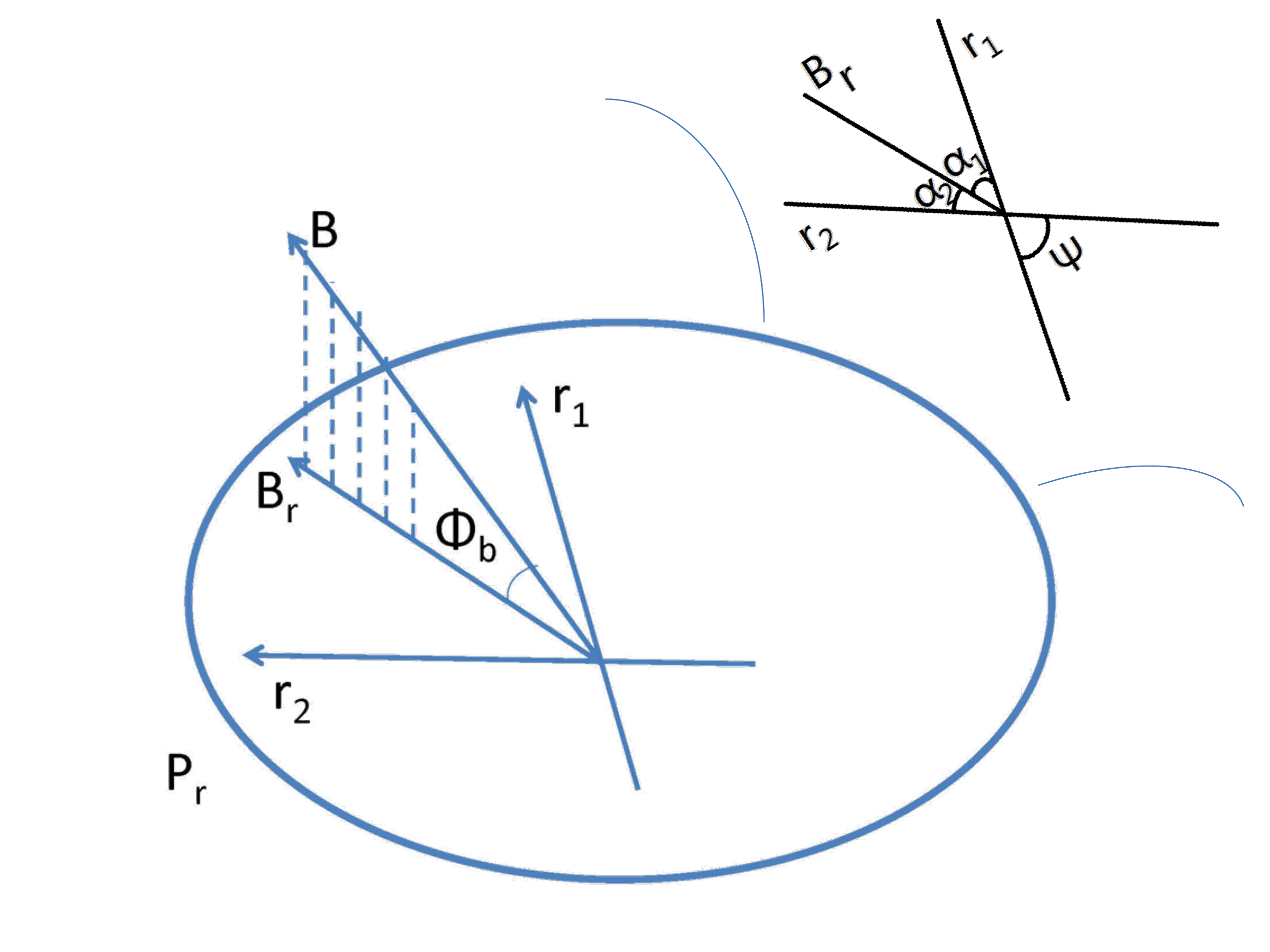}\label{fig6b}}
\caption{
(a) Simplification for a binary system. A and B: two pumping sources; C: the medium where the alignment happens. As demonstrated in \S 4, if the medium is enough distant from the pumping source, the radiation spheres can be simplified as two points.
(b) Geometry of a binary radiation system. $r_1$, $r_2$: radiation lines from the two sources; $P_r$: the plane of the binary system; $B$: the direction of magnetic field; $B_r$: the projection of B on the plane $P_r$; $\phi_b$: the inclination angle of magnetic field (the angle between the radiation plane and magnetic field). The geometric relation in the $P_r$ plane is further illustrated in the upper-right plot. $\alpha_1\And\alpha_2$: the angles between $B_r$ and two radiation lines $r_1$, $r_2$; $\psi$: the angle between $r_1$ and $r_2$.}
\end{figure*}

As demonstrated in \S 4, the pumping by a beam of light is the asymptotic limit of the pumping by a narrow cone of light. Hence, for the sake of simplicity, the binaries are treated as two point sources in the case that the medium considered is distant enough, as shown in Fig.~\ref{fig6a}. It is generally believed that circumbinary gaseous disks are geometrically coplanar with the binaries \citep[see][]{1994ApJ...421..651A}. The geometry of the binary system is illustrated in Fig.~\ref{fig6b}.

From Fig.~\ref{fig6b}, we know $\alpha_1=\Psi-\alpha_2$. Thus the angles between magnetic field and two radiation lines can be represented as $(\cos\alpha_{1}\cos\phi)$ and $(\cos(\Psi-\alpha_{1})\cos\phi)$, respectively. We define $r$ as the intensity ratio of the radiations from $r_1$ and $r_2$:
\begin{equation}
r=I_{r_1}/I_{r_2},
\end{equation}

\begin{figure*}
\centering
\subfigure[]{
\includegraphics[width=0.45\columnwidth,
 height=0.28\textheight]{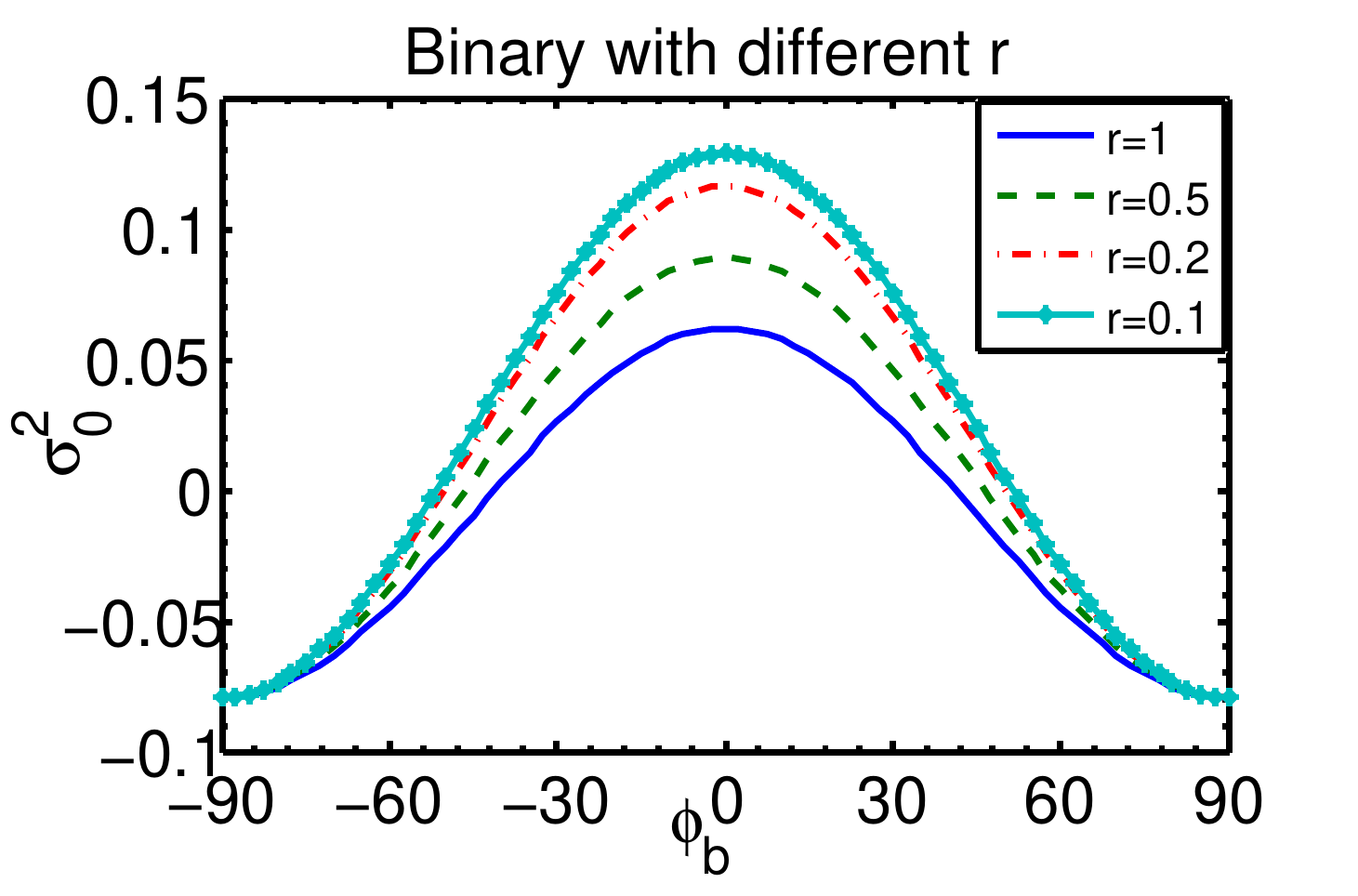}\label{fig7a}}
\subfigure[]{
\includegraphics[width=0.45\columnwidth,
 height=0.28\textheight]{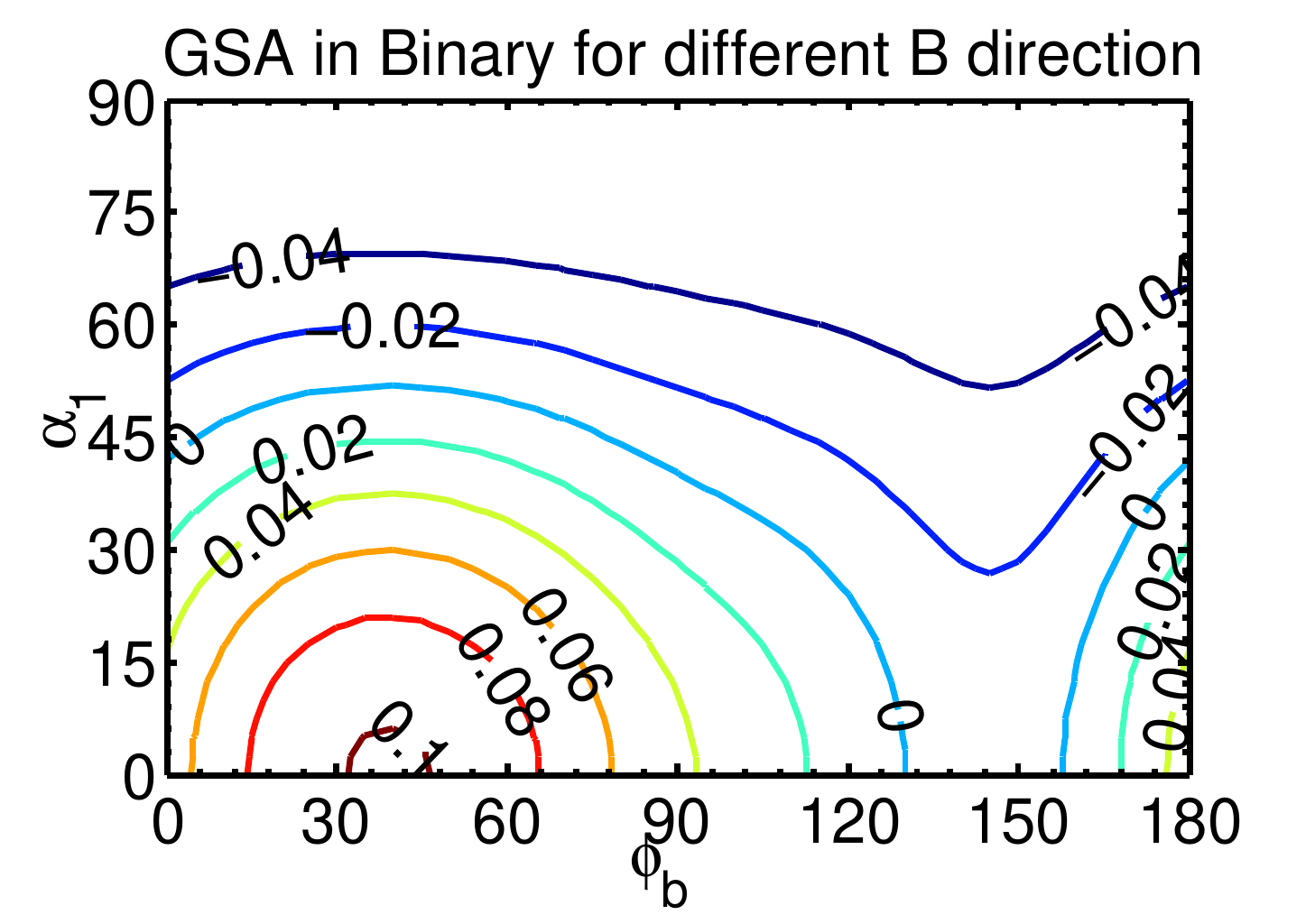}\label{fig7b}}
\caption{GSA of SII in Binary system.
(a) Alignment parameter vs. $\phi_b$ for different intensity ratio $r$ with $\alpha_1=\pi/8$; $\psi=\pi/2$.
(b) Iso-contour plot for the polarization on $\phi_b$ and $\alpha_1$ in a binary system where $\psi=\pi/3$ and $r=0.5$.
}
\end{figure*}
\begin{figure*}
\centering
 \subfigure[]{
\includegraphics[width=0.45\columnwidth,
 height=0.28\textheight]{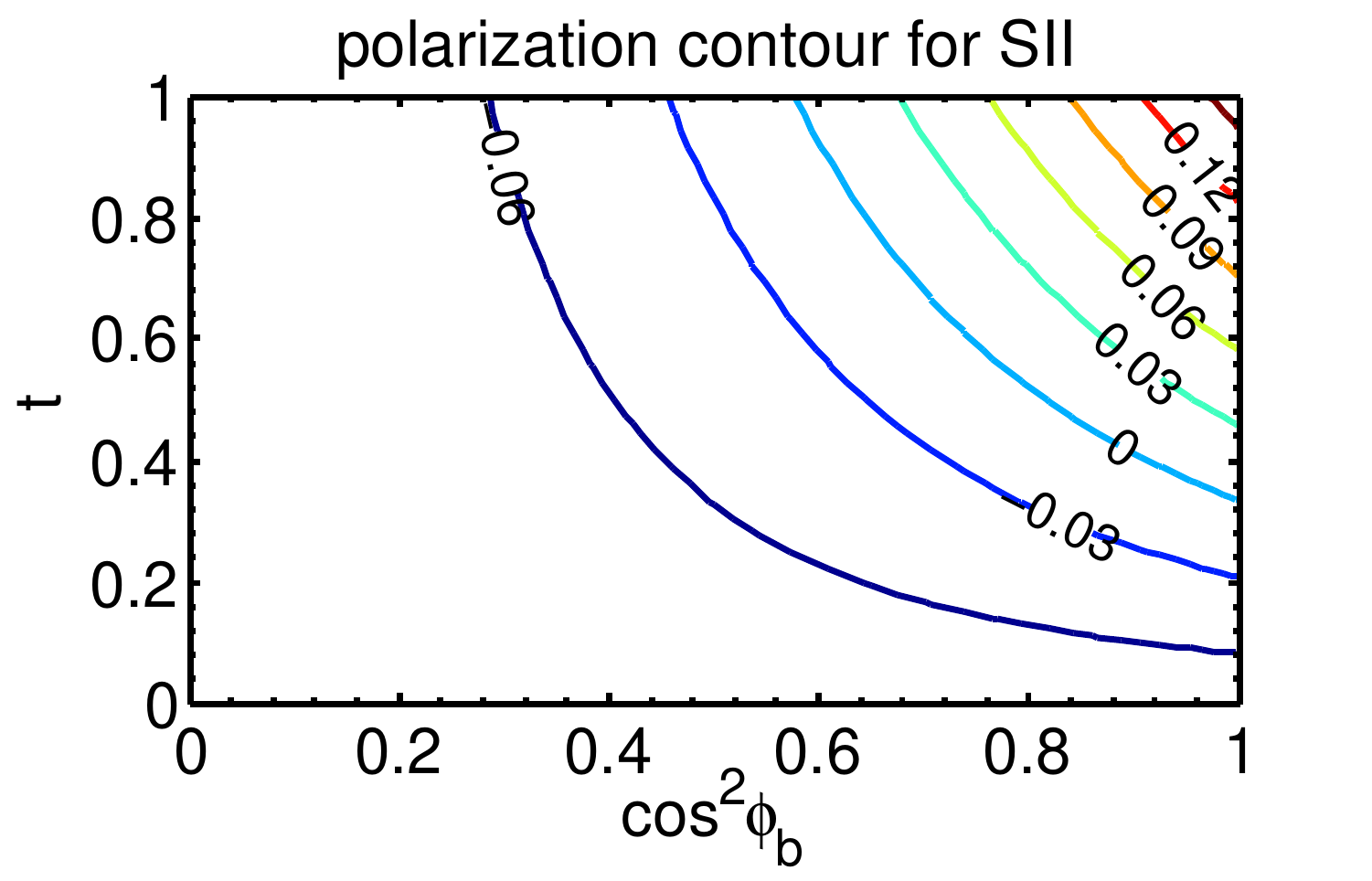}\label{fig8a}}
 \subfigure[]{
\includegraphics[width=0.45\columnwidth,
 height=0.28\textheight]{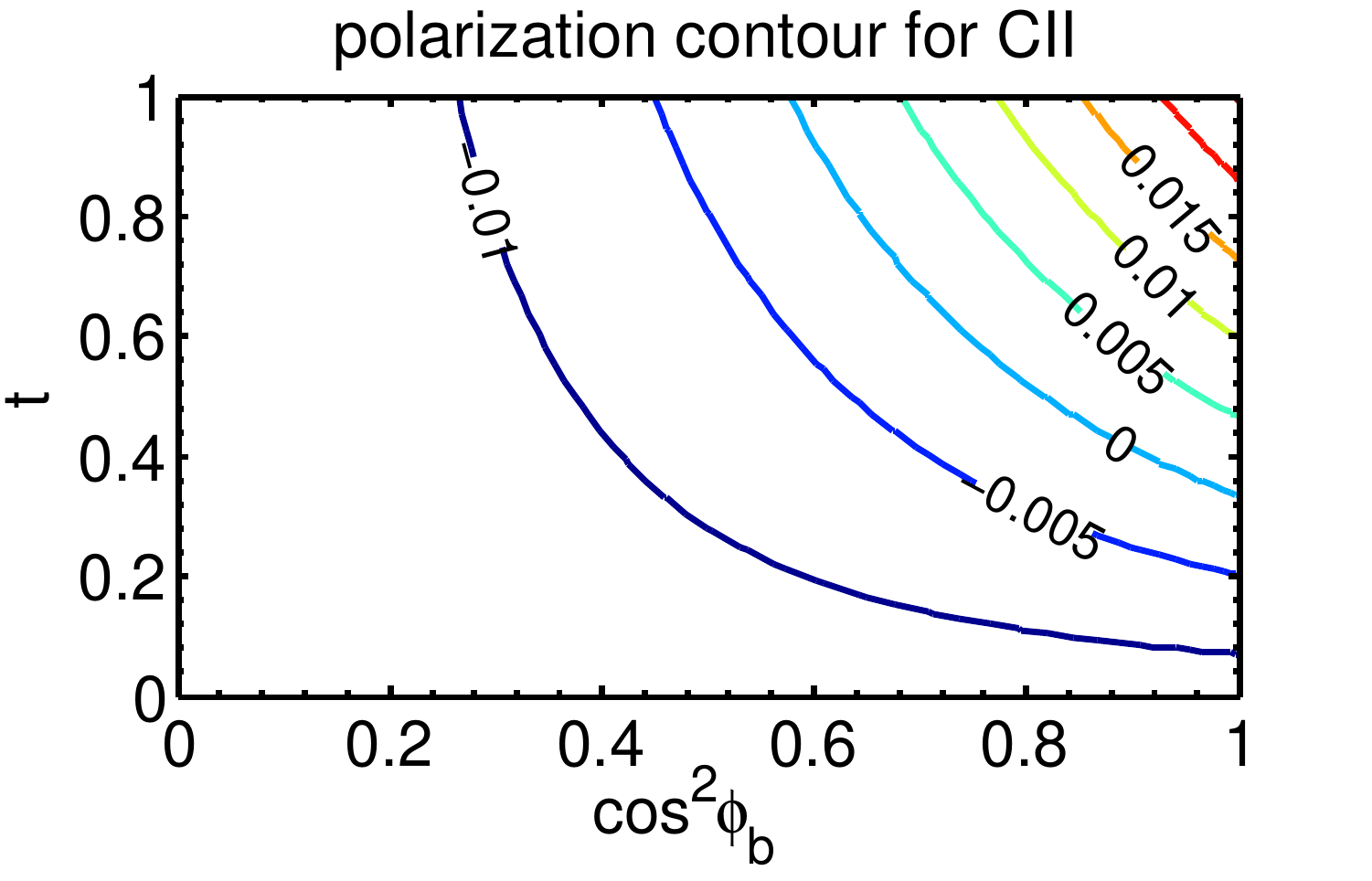}\label{fig8b}}
\caption{Iso-contour plots for the polarization in Binary system on $t$ and $\cos^2\phi_b$ in the use of (a) SII absorption line and (b) CII absorption line, respectively.
}
\end{figure*}

The radiation tensor is obtained by summing up the contribution from the two radiations:
\begin{equation}\label{binarytensor}
\bar{J}_{0}^{2}=\frac{1}{2\sqrt{6}} \left(3 \left(\frac{\cos^{2}\alpha_{1}+r\cos^{2}(\Psi-\alpha_{1})}{1+r}\cos^{2}\phi_{b} \right) -1 \right).
\end{equation}

Obviously, once the ratio $r$ and the angle $\Psi$ are known, the degree of polarization is solely determined by the direction of magnetic field ($\phi_b$, $\alpha_1$). In addition, we define
\begin{equation}\label{tdef}
t=\frac{\cos^{2}\alpha_{1}+r\cos^{2}(\Psi-\alpha_{1})}{1+r}
\end{equation}
to measure the anisotropy of the binary system. Apparently, when $r\rightarrow 0$ or $\infty$, the system falls back to the case of a beam of light.

Infixing Eq.~\eqref{binarytensor} into Eq.~\eqref{groundoccu}, we obtain the alignment parameter of SII in a binary system:
\begin{equation}\label{binarydens}
\sigma_{0}^{2}(J_l)=\frac{0.4199-1.2596t\cos^{2}\phi_{b}}{0.1490t\cos^{2}\phi_{b}-5.2868}.
\end{equation}

Substituting Eq.~\eqref{binarydens} into Eq.~\eqref{ptau}, we get the polarization of SII absorption line in a binary system:
\begin{equation}
\frac{P}{\tau}=\frac{(0.6299-1.8894t\cos^{2}\phi_b)\sin^{2}\theta \omega_{J_lJ_u}^2}{0.2107t\cos^{2}\phi_{b}-7.4767+( 0.4199-1.2596t\cos^{2}\phi_{b})(1-1.5\sin^{2}\theta)\omega_{J_lJ_u}^2}.
\end{equation}

The results for binary systems with different ratio $r$ are demonstrated in Fig.~\ref{fig7a}, showing that the closer the intensity of the secondary is to that of the primary, the smaller the alignment is. This is because the binary system becomes more isotropic as the radiation from the secondary gets comparable to that from the primary. As an example, a binary system with $\psi=\pi/3$ and $r=0.5$ is discussed in Fig.~\ref{fig7b} for the polarization with varying magnetic field ($\phi_b$, $\alpha_1$). The polarization is apparently dependent on the direction of magnetic field.

The iso-contour plots for polarization with varying binary parameter ($t$) and magnetic field inclination ($\phi_b$) are presented in Fig. 8. The polarization crosses zero when $\phi_b=\arccos\frac{1}{\sqrt{3t}}$. It is a general feature in binary systems for any atomic species that the polarization of absorption line flips between being parallel and perpendicular to magnetic field when the inclination angle of magnetic field $\phi_b$ is $\arccos\frac{1}{\sqrt{3t}}$. {\em Therefore, 2D magnetic field in binary systems can be directly determined by the direction of polarization with a $90^\circ$ degeneracy, whereas 3D magnetic field can be obtained if the degree of polarization is observed.}

\section{GSA with disc shape radiation field}

In this section, we apply GSA to the disc shape radiation field, and discuss the alignment in the Local Interstellar Medium(LISM) as an example in \S 6.1.

\begin{figure*}
\centering
\subfigure[]{
\includegraphics[width=0.45\columnwidth,
 height=0.28\textheight]{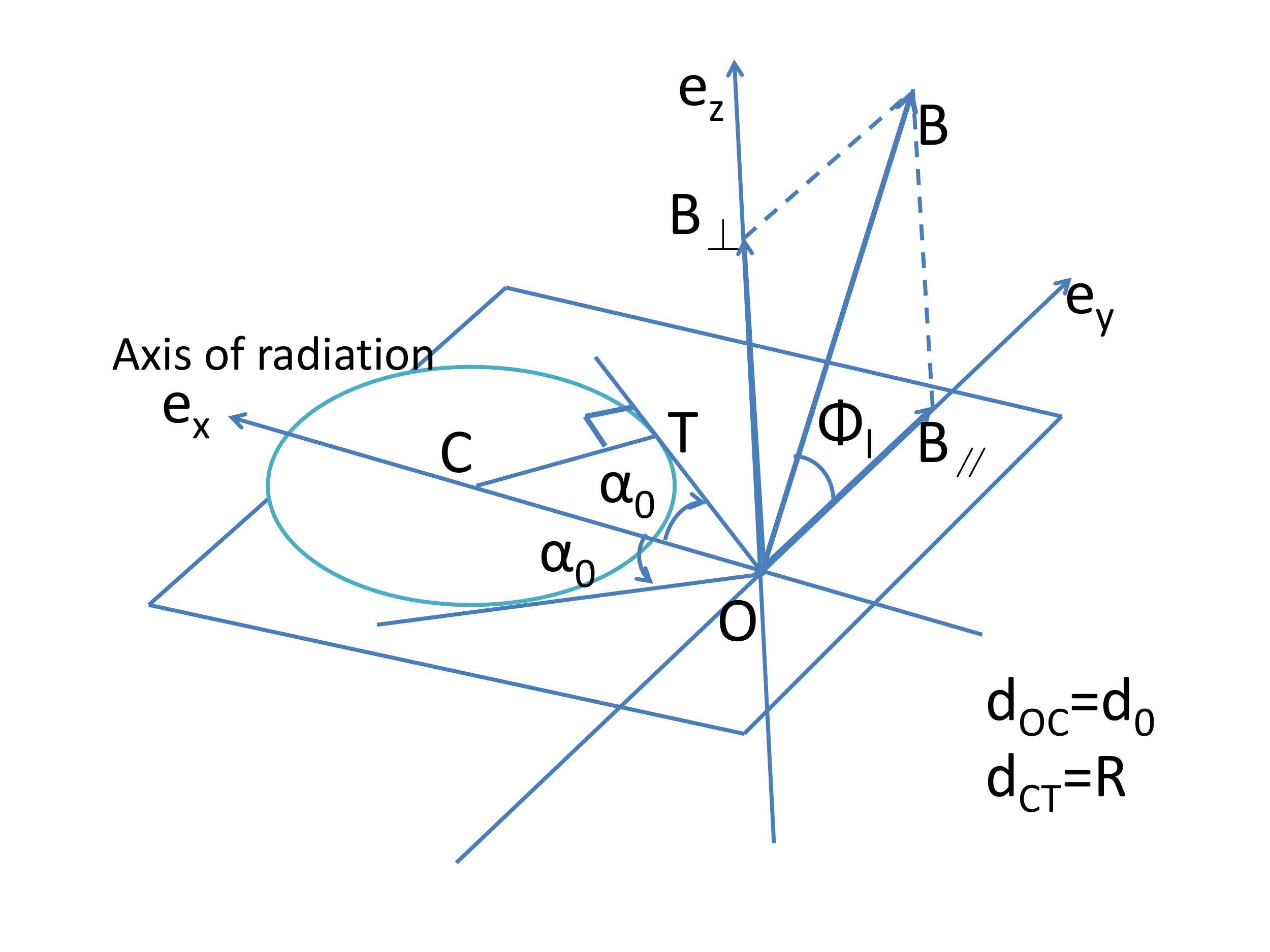}\label{fig9a}}
 \subfigure[]{
\includegraphics[width=0.45\columnwidth,
 height=0.28\textheight]{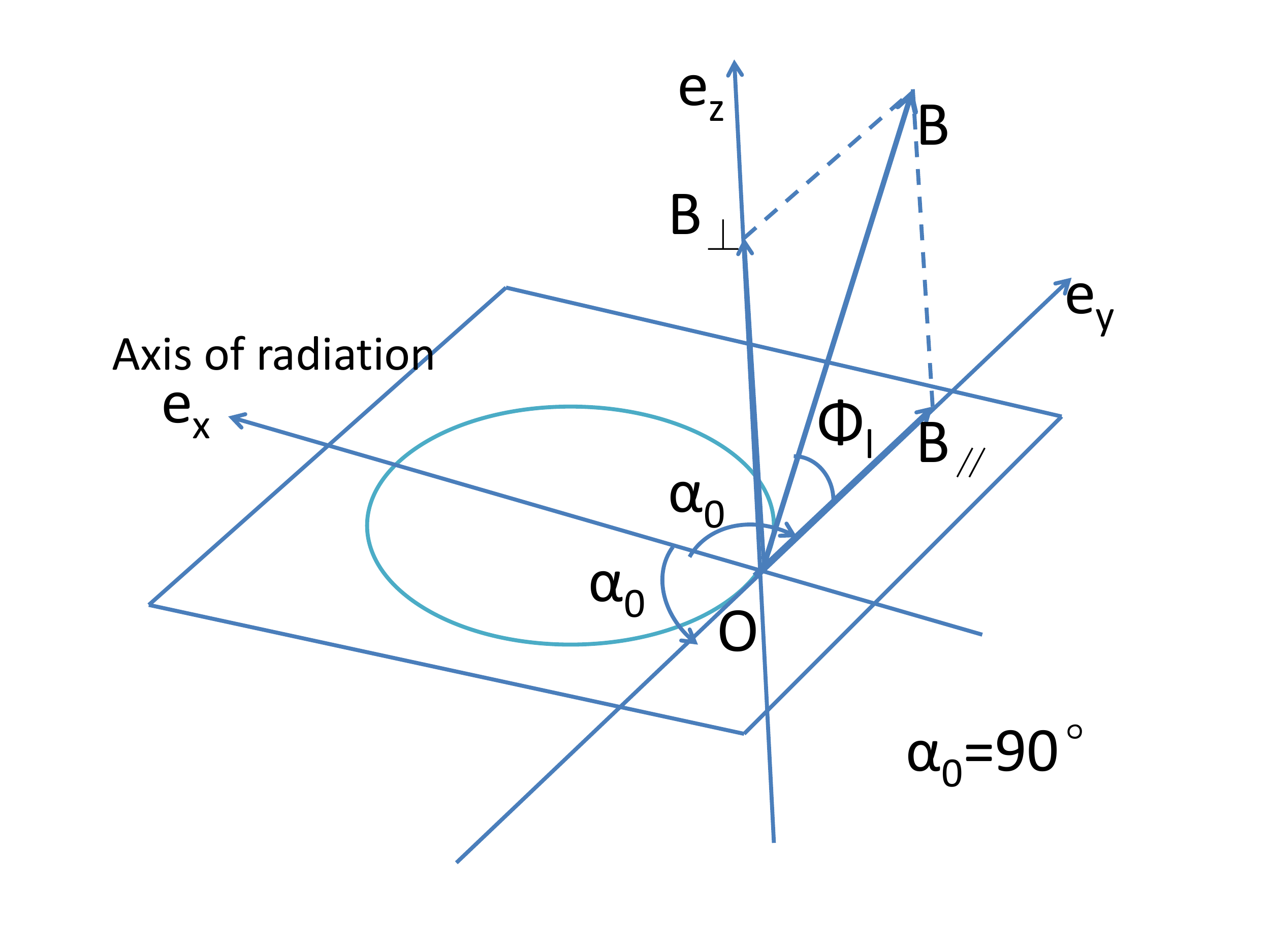}\label{fig9b}}
\caption{(a) Geometry for GSA with the disc shape radiation. The whole circle on the left is the pumping source. O: the diffuse medium in consideration; C: the center of the source. $B$: magnetic field. $B_{\|}\And B_{\perp}$: the projections of magnetic field parallel and perpendicular to the disc. OT: the tangential line; $e_x$: the symmetric axis of the radiation disc. $\alpha_0$: the angle between CO and OT. $\phi_l$: the inclination angle of magnetic field (the angle between $B$ and $B_{\|}$). $R$: radius of the disc shape radiation. $d_0$: distance from the center of the disc to the diffuse medium.
(b) Geometry for GSA in the LISM. Diffuse medium O is on the edge of the disc radiation field, which means $\alpha_0=90^{\circ}$.
}
\end{figure*}

The disc shape radiation field is illustrated in Fig.~\ref{fig9a}. The intensity of the pumping source is evenly distributed, and $2\alpha_0=2\arcsin (R/d_0)$ is the flare angle of the radiation field. The radius of the disc is much larger than the thickness of the disc. Hence, the thickness of the radiation source can be neglected and thin disc approximation can be applied. It is assumed that magnetic field has two components: the one along the circumference of the disc and that perpendicular to the disc, hereby $B_{\|}$ and $B_{\perp}$, which means the projection of magnetic field on the plane of disc is tangential to the disc, e.g., the case of galactic magnetic field in Spiral Arm.

The radiation tensor can be obtained by integrating all the radiation from the disc:
\begin{equation}\label{disctensor}
\bar{J}_{0}^{2}=\frac{1}{2\sqrt{6}} \left(\frac{3}{2}\cos^{2}\phi_l f_l(\alpha_0)-1 \right),
\end{equation}
where
\begin{equation}\label{fldef}
f_l(x)=1-\frac{\sin{2x}}{2x},\phi_l=\arctan{\frac{B_{\perp}}{B_{\|}}}.
\end{equation}

The alignment parameter of SII in disc shape radiation field is obtained by infixing Eq.~\eqref{disctensor} into Eq.~\eqref{groundoccu}:
\begin{equation}\label{discdens}
\sigma_{0}^{2}(J_l)=\frac{6.2982\cos^{2}\phi_{l}f_l(\alpha_0)-4.1988}{52.8680-0.7448\cos^{2}\phi_{l}f_l(\alpha_0)}
\end{equation}

Substituting Eq.~\eqref{discdens} into Eq.~\eqref{ptau}, we obtain the polarization of SII absorption line in disc shape radiation induced by GSA:
\begin{equation}\label{discpolar}
\frac{P}{\tau}=\frac{(-9.4473\cos^{2}\phi_{l}f_l(\alpha_0)+6.2982)\sin^{2}\theta \omega_{J_lJ_u}^2}{74.7666-1.0533\cos^{2}\phi_{l}f_l(\alpha_0)+(4.1988-6.2982\cos^{2}\phi_{l}f_l(\alpha_0))(1-1.5\sin^{2}\theta)\omega_{J_lJ_u}^2}.
\end{equation}

The radiation tensor $\bar{J}_0^2$ equals to zero when the inclination angle of magnetic field $\phi_l$ is $\arccos\sqrt\frac{2}{3f_l(\alpha_0)}$, as demonstrated in Eq.\eqref{disctensor}. This leads to the sign reversal of the polarization of absorption line in disc shape radiation field at $\arccos\sqrt\frac{2}{3f_l(\alpha_0)}$. As a result, the polarization switches between being parallel and perpendicular to magnetic field when $\phi_l=\arccos\sqrt\frac{2}{3f_l(\alpha_0)}$.

\begin{figure*}
\centering
\subfigure[]{
\includegraphics[width=0.45\columnwidth,
 height=0.28\textheight]{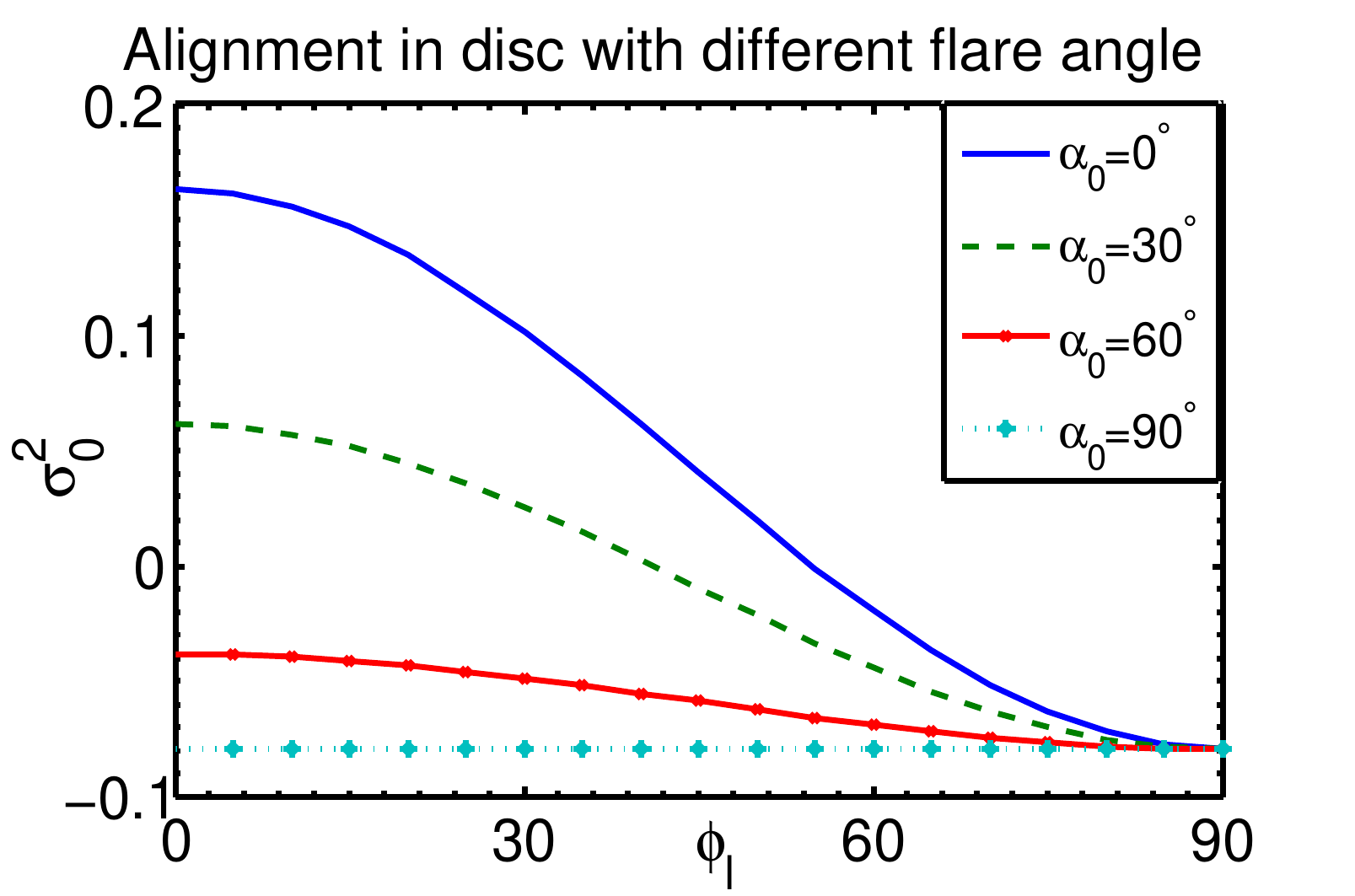}\label{fig10a}}
\subfigure[]{
\includegraphics[width=0.45\columnwidth,
 height=0.28\textheight]{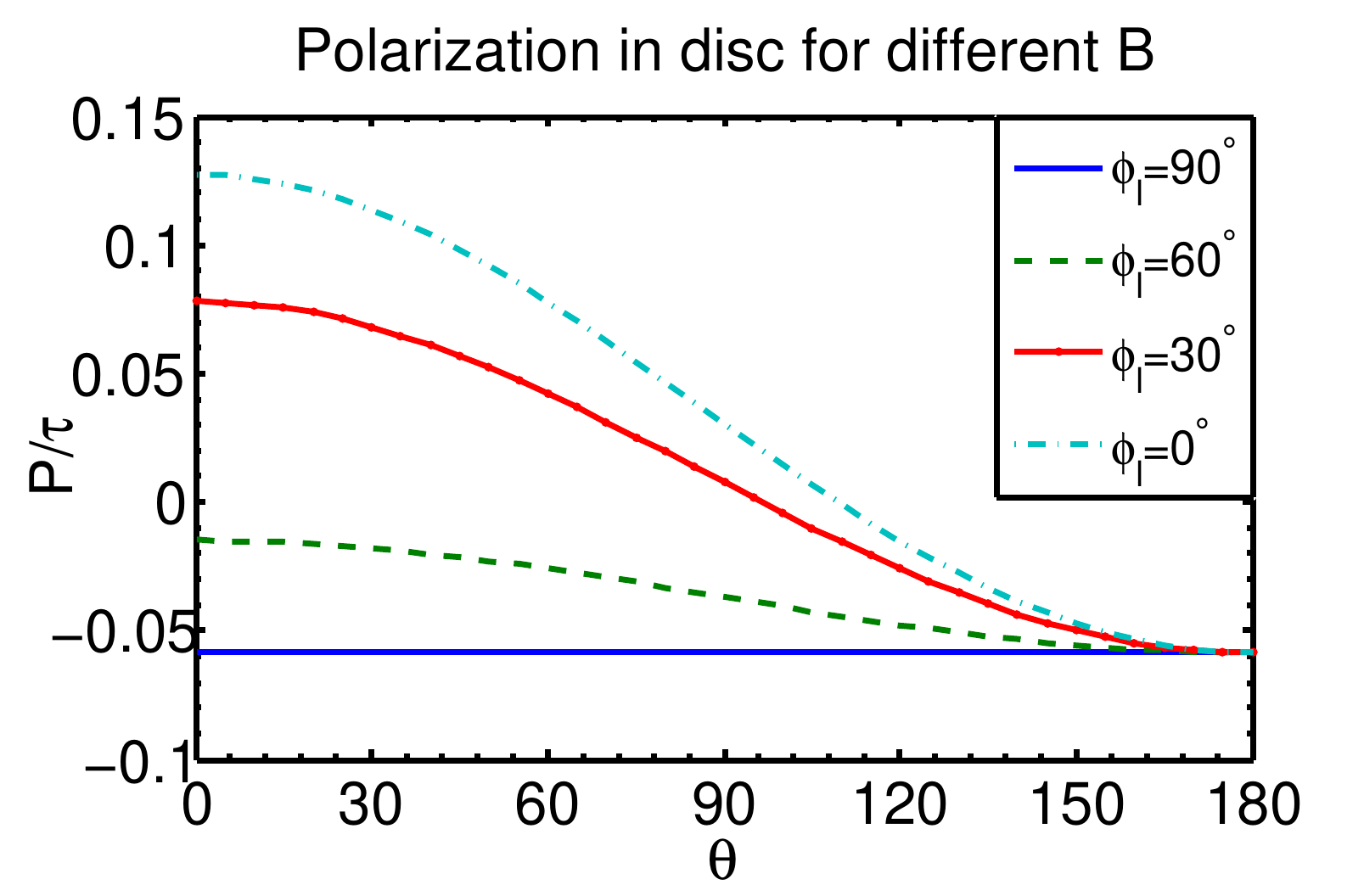}\label{fig10b}}
\caption{ GSA with disc shape radiation field. The physical parameters defined in Fig.~\ref{fig9a} and Fig.~\ref{fig9b}. (a) The alignment of SII with different flare angle. The line with $\alpha_0=0$ corresponds to the case
 of the alignment in a beam of light. (b) Polarization of SII absorption line with different magnetic inclination.}
\end{figure*}

Fig.~\ref{fig10a} depicts the change of the alignment with the disc flare angle ($2\alpha_0$). Smaller flare angle means longer distance from the medium to the pumping source. Obviously, the alignment in disc shape radiation field approaches to that in a beam of light \citep[see][]{YLfine} as $\alpha_{0}\rightarrow 0$. Fig.~\ref{fig10b} shows the polarization with different magnetic inclination $\phi_l$, indicating that more proportion of $B_{\perp}$ reduces the degree of polarization. In conclusion, the direction of magnetic field in disc shape radiation field can be determined by the observation of the polarization induced by GSA.

\begin{figure*}
\centering
\subfigure[]{
\includegraphics[width=0.45\columnwidth,
 height=0.28\textheight]{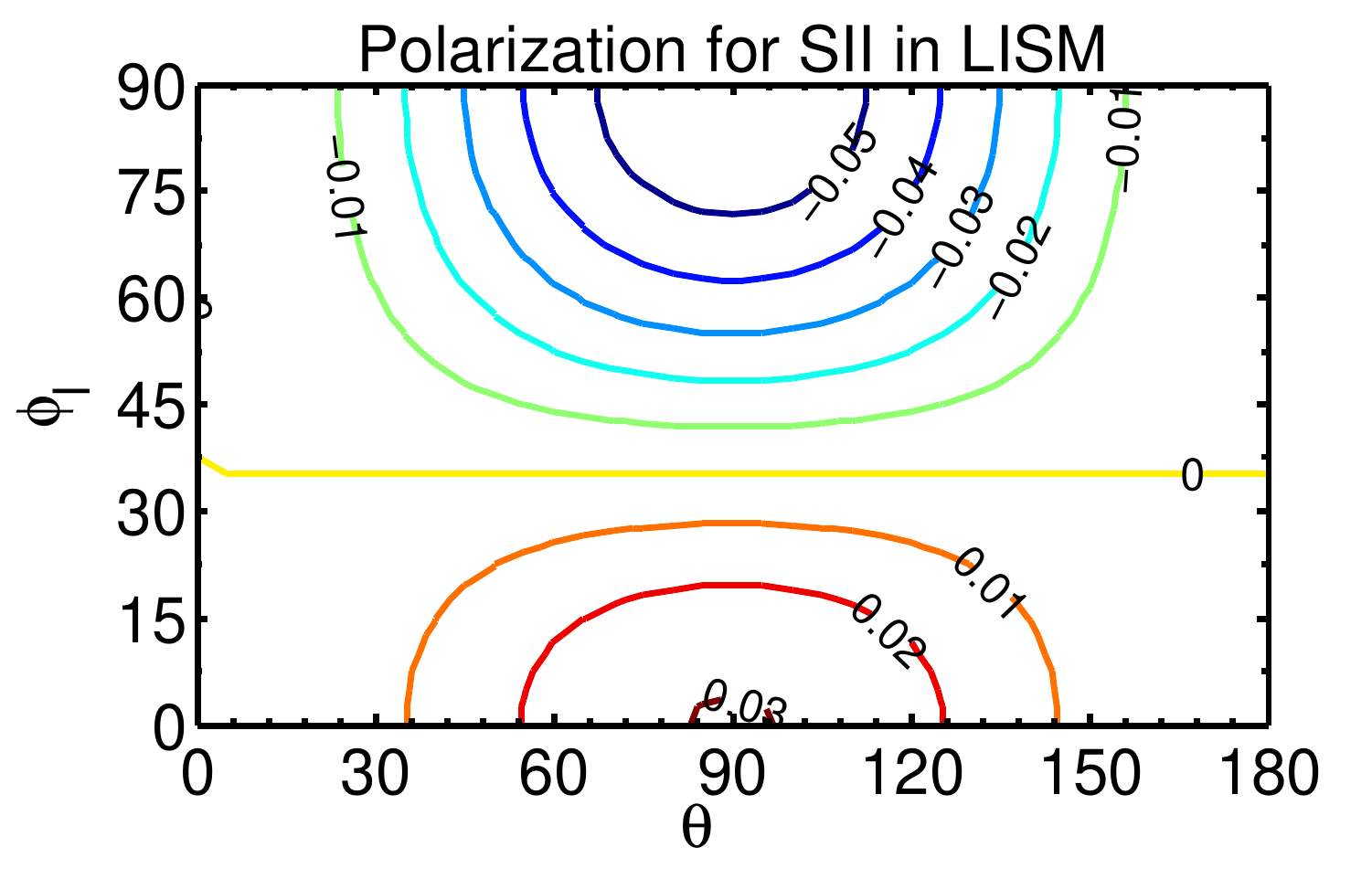}\label{fig11a}}
 \subfigure[]{
\includegraphics[width=0.45\columnwidth,
 height=0.28\textheight]{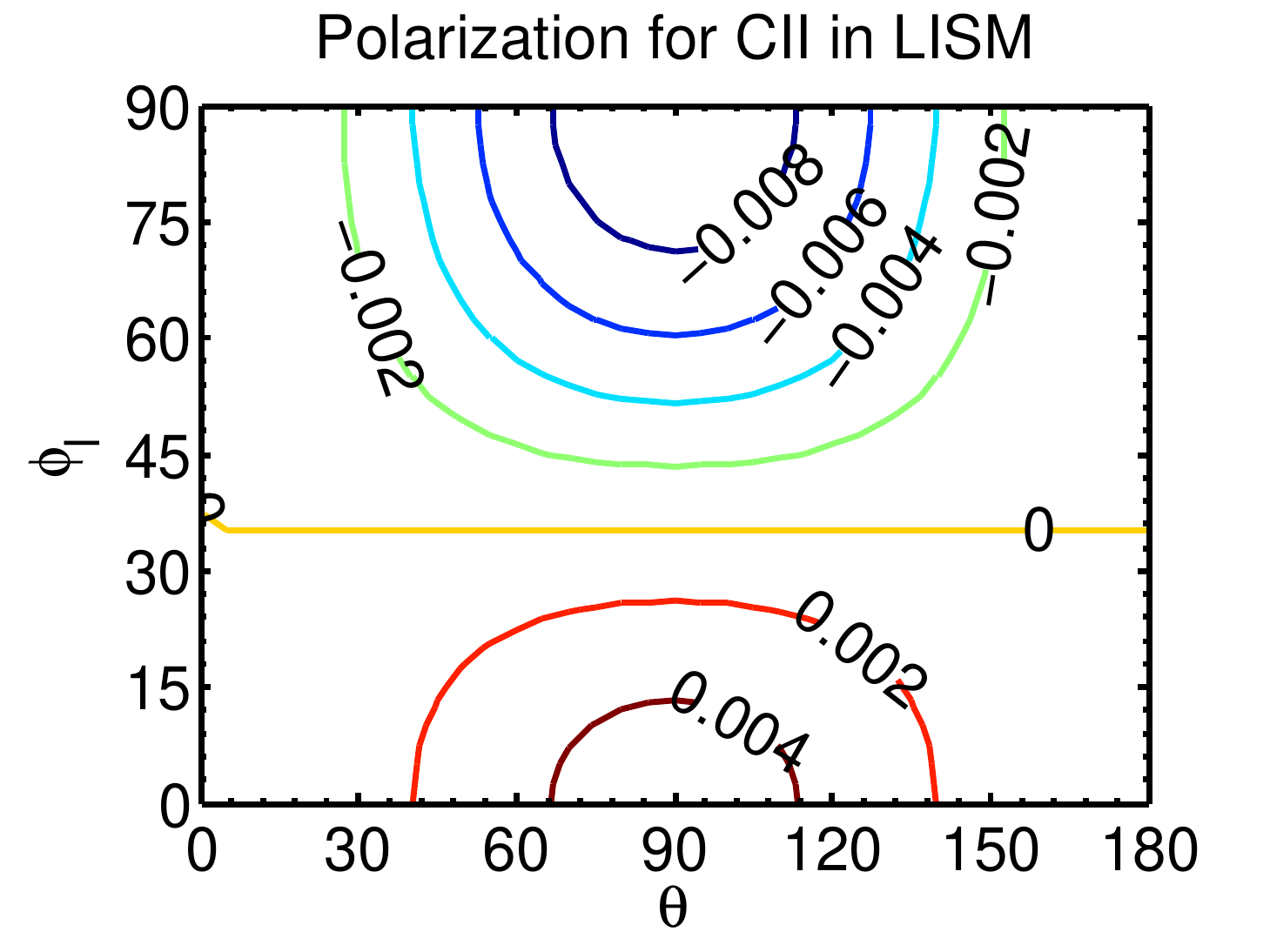}\label{fig11b}}
\caption{Iso-contour plots for the polarization with varying $\phi_l$ and $\theta$ for (a) SII absorption line and (b) CII absorption line, respectively, in the Local ISM.}
\end{figure*}
\subsection{GSA in the LISM}

The radiation in the LISM is concentrated in a thin disc (e.g., see \citealt{1982A&A...105..372M}). Inside the LISM, the flare angle is
\begin{equation}\label{lismflare}
2\alpha_0=180^{\circ}.
\end{equation}

Substituting Eq.~\eqref{lismflare} into Eq.~\eqref{disctensor}, Eq.~\eqref{discdens}, and Eq.~\eqref{discpolar}, we obtain the radiation tensor for the LISM:
\begin{equation}\label{lismtensor}
\bar{J}_{0}^{2}=\frac{1}{4\sqrt{6}} \left(3\cos^{2}\phi_l-2 \right),
\end{equation}
and accordingly the alignment parameter of SII in the LISM:
\begin{equation}\label{lismdens}
\sigma_{0}^{2}(J_l)=\frac{6.2982\cos^{2}\phi_{l}-4.1988}{52.8680-0.7448\cos^{2}\phi_{l}},
\end{equation}
and the polarization of SII absorption line in the LISM:
\begin{equation}\label{lismcpolar}
\frac{P}{\tau}=\frac{(-9.4473\cos^{2}\phi_{l}+6.2982)\sin^{2}\theta \omega_{J_lJ_u}^2}{74.7666-1.0533\cos^{2}\phi_{l}+(4.1988-6.2982\cos^{2}\phi_{l})(1-1.5\sin^{2}\theta)\omega_{J_lJ_u}^2}.
\end{equation}

The iso-contour plot of polarization for varying direction of line of sight ($\theta$) and magnetic field inclination ($\phi_l$) is provided in Fig.11. The polarization of absorption line in the LISM switches between being parallel and perpendicular to the direction of magnetic field when $\phi_l$ is $35.3^{\circ}$. {\em Hence, 2D magnetic field can be directly determined by the direction of polarization with a $90^{\circ}$ degeneracy. The 3D information of magnetic field in the LISM can be obtained if the degree of polarization is observed.}

\begin{figure*}
\centering
\subfigure[]{
\includegraphics[width=0.33\columnwidth,
 height=0.22\textheight]{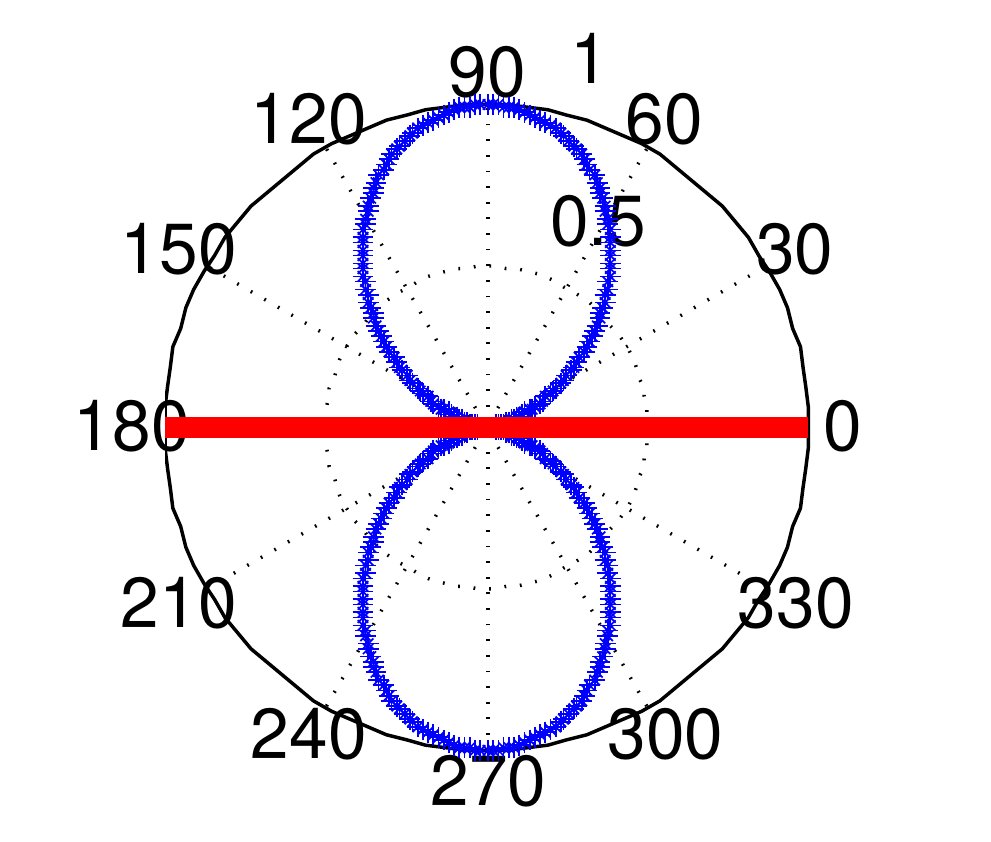}\label{fig12a}}
\subfigure[]{
\includegraphics[width=0.33\columnwidth,
 height=0.19\textheight]{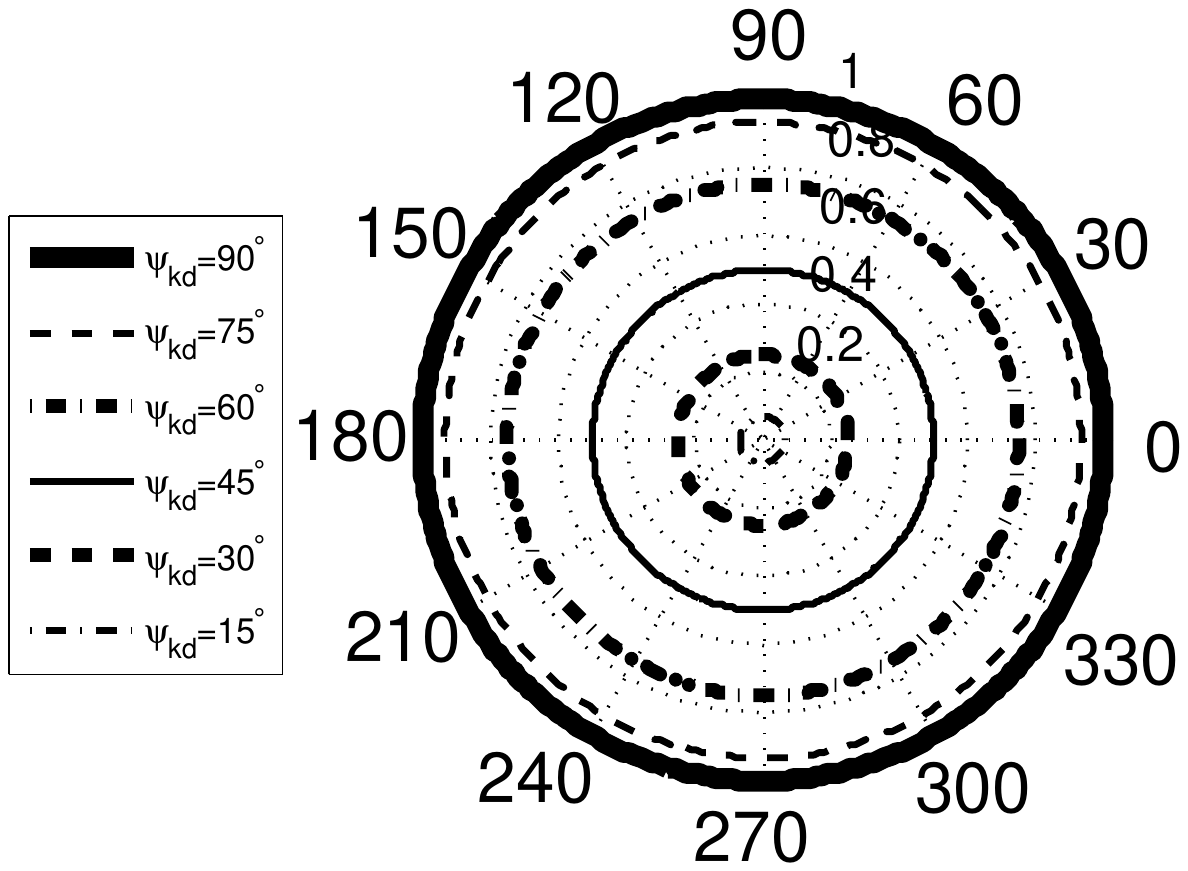}\label{fig12b}}
 \subfigure[]{
 \includegraphics[width=0.3\columnwidth,
 height=0.23\textheight]{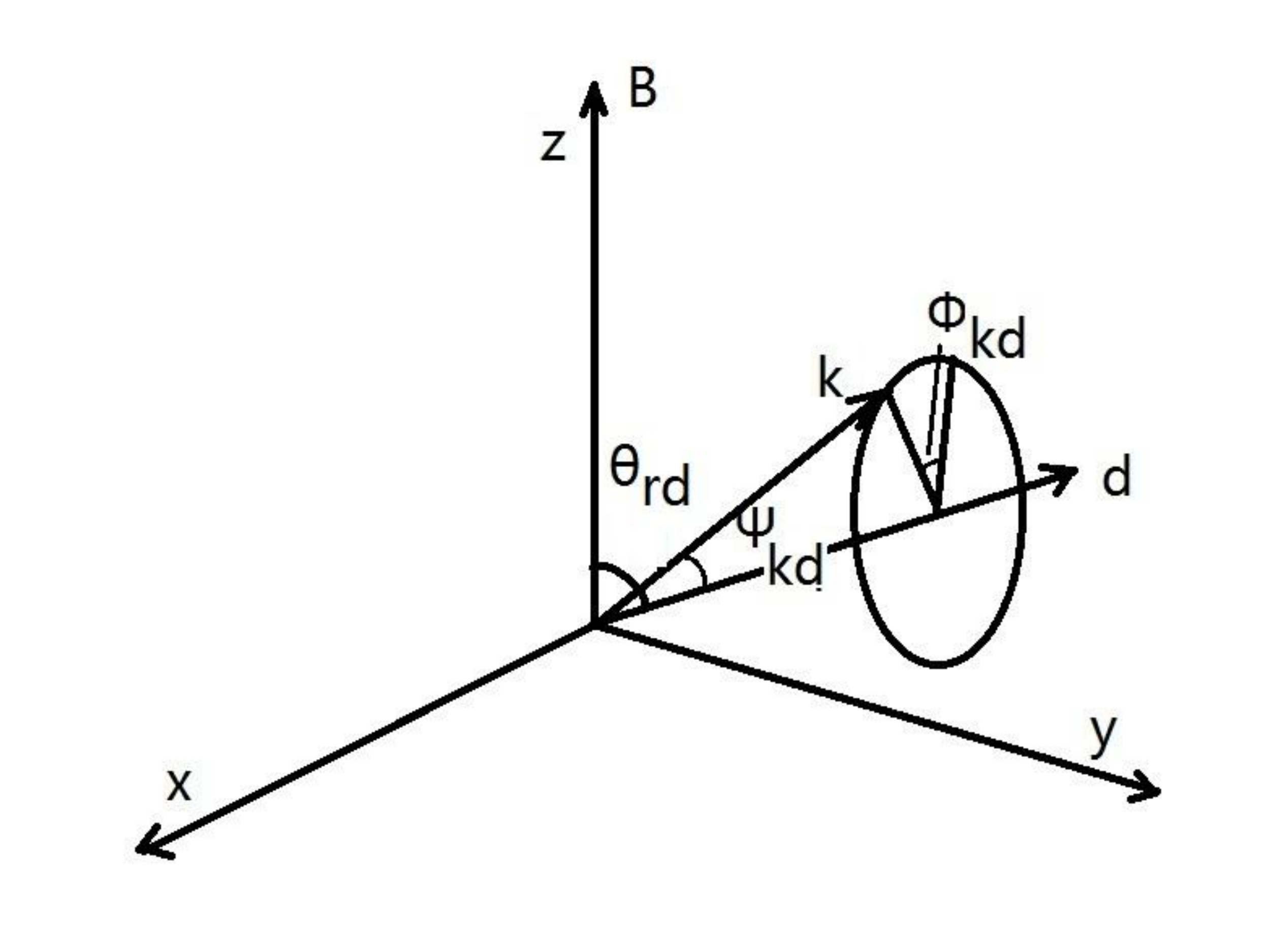}\label{fig12c}}
\caption{Dipole radiation field.
(a)$\And$(b) Geometry of the dipole radiation field in the direction perpendicular and parallel to the symmetrical axis, respectively.
(c) Coordinate system of the dipole radiation field. $d$: the dipole axis. $k$: the radiation direction. $B$: magnetic field. $\theta_{rd}$: the angle between the dipole axis $d$ and magnetic field $B$. ($\psi_{kd},$ $\phi_{kd}$): angle coordinate of $k$ in the dipole radiation system.}
\end{figure*}

\begin{figure*}
\centering
\subfigure[]{
\includegraphics[width=0.33\columnwidth,
 height=0.22\textheight]{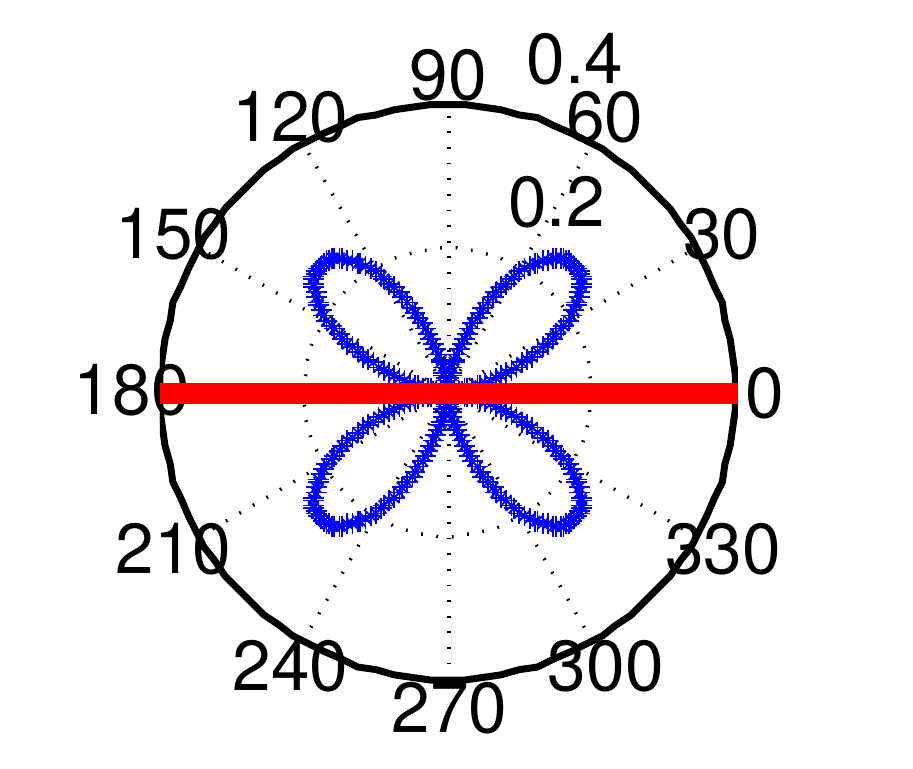}\label{fig13a}}
\subfigure[]{
\includegraphics[width=0.33\columnwidth,
 height=0.19\textheight]{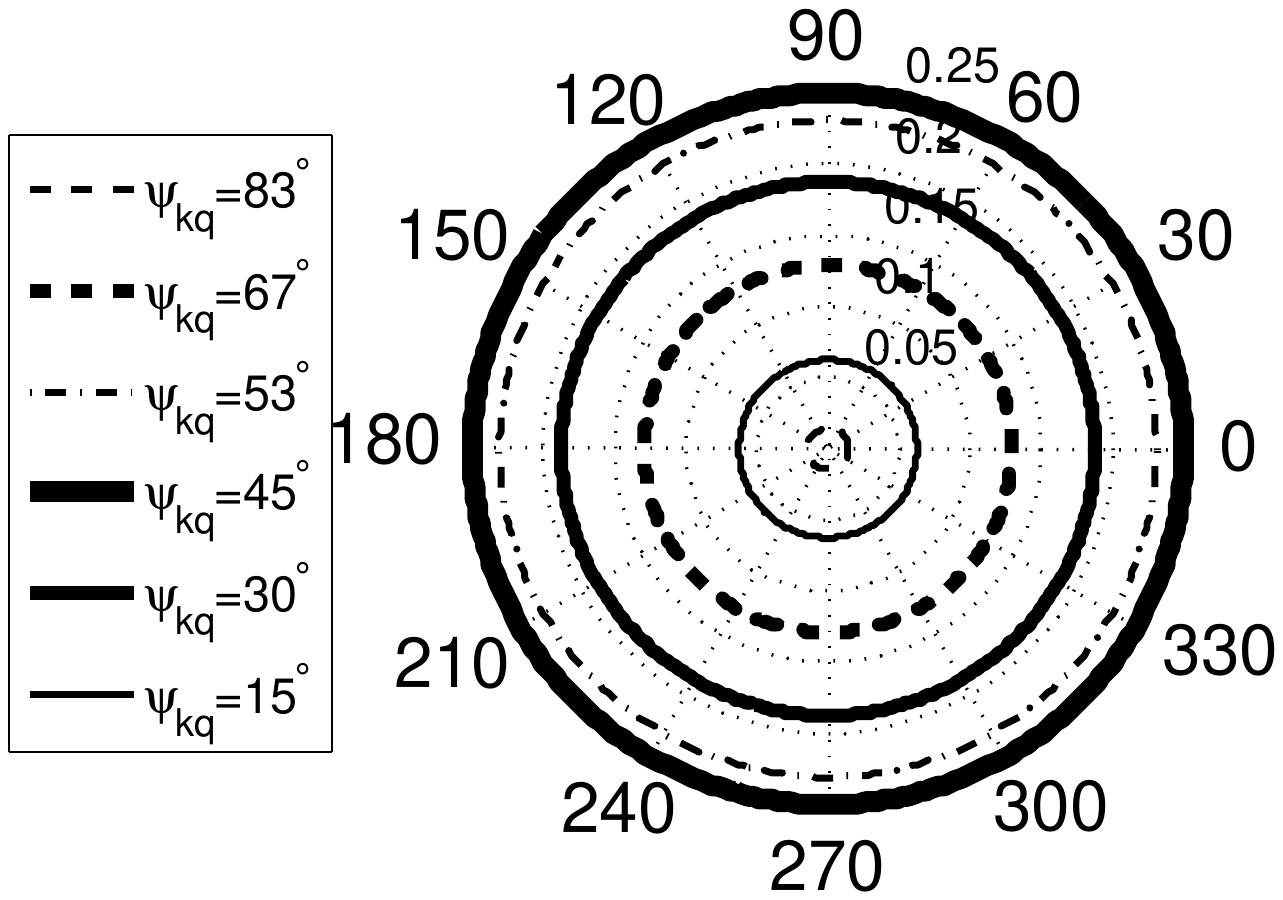}\label{fig13b}}
 \subfigure[]{
 \includegraphics[width=0.3\columnwidth,
 height=0.23\textheight]{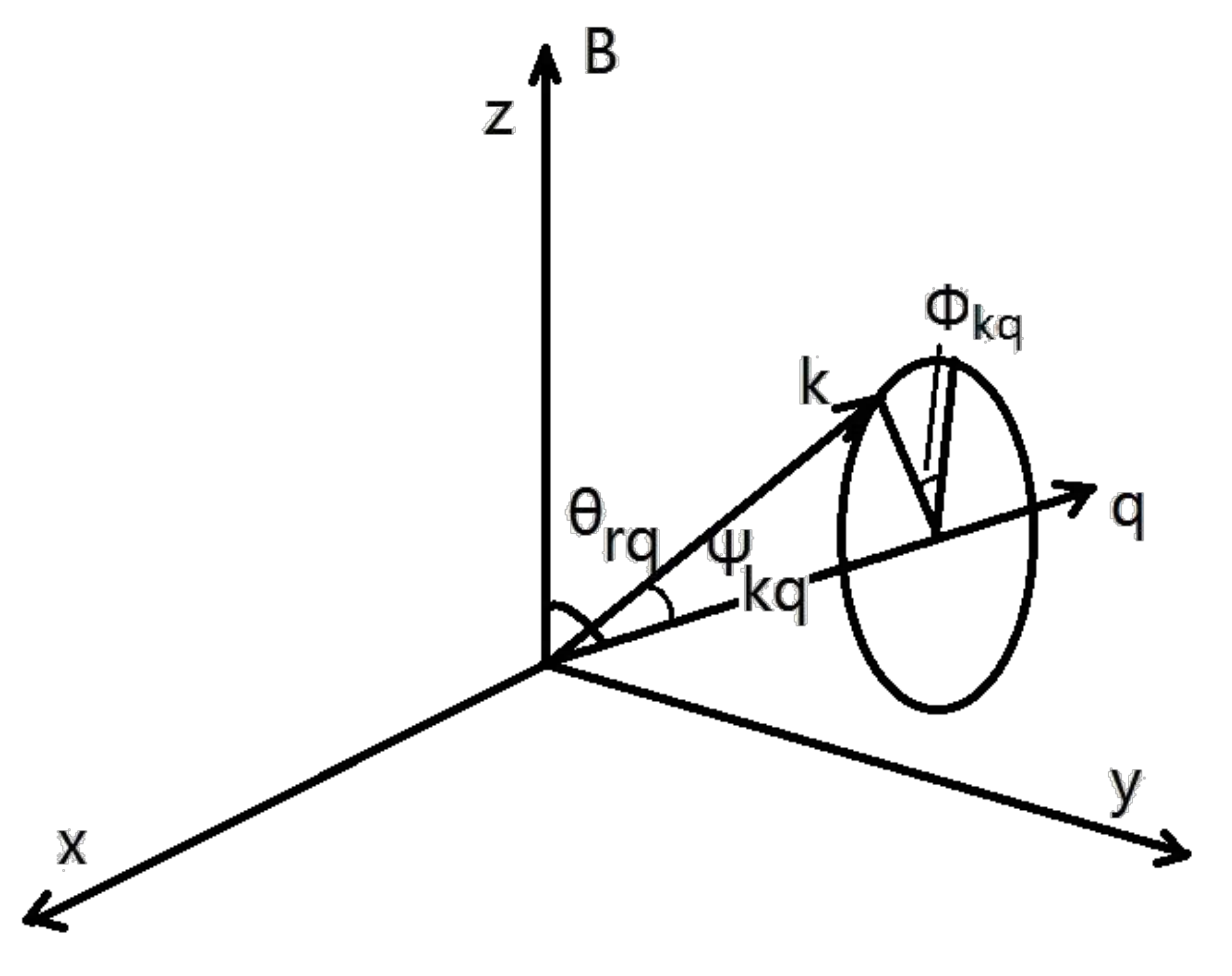}\label{fig13c}}
\caption{Quadrupole radiation field.
(a)$\And$(b) Geometry of the quadrupole radiation field in the direction perpendicular and parallel to the symmetrical axis, respectively.
(c) Coordinate system of the quadrupole radiation field. $q$: the quadrupole axis. $k$: the radiation direction. $B$: magnetic field. $\theta_{rq}$: the angle between the quadrupole axis $d$ and magnetic field $B$. ($\psi_{kq},$ $\phi_{kq}$): angle coordinate of $k$ in the quadrupole radiation system.}
\end{figure*}

\section{GSA with General Radiation field}

It has been discussed in the previous sections that GSA can be applied to trace magnetic field where the specific pumping sources of the radiation field are identified. However, sources for radiation field may not be easily identified in some circumstances. In this case, the method of multipole expansion can be adopted. That is, to compute the alignment in multipole components and then add them up.

\begin{figure}
\centering
\subfigure[]{
\includegraphics[width=0.45\columnwidth,
 height=0.28\textheight]{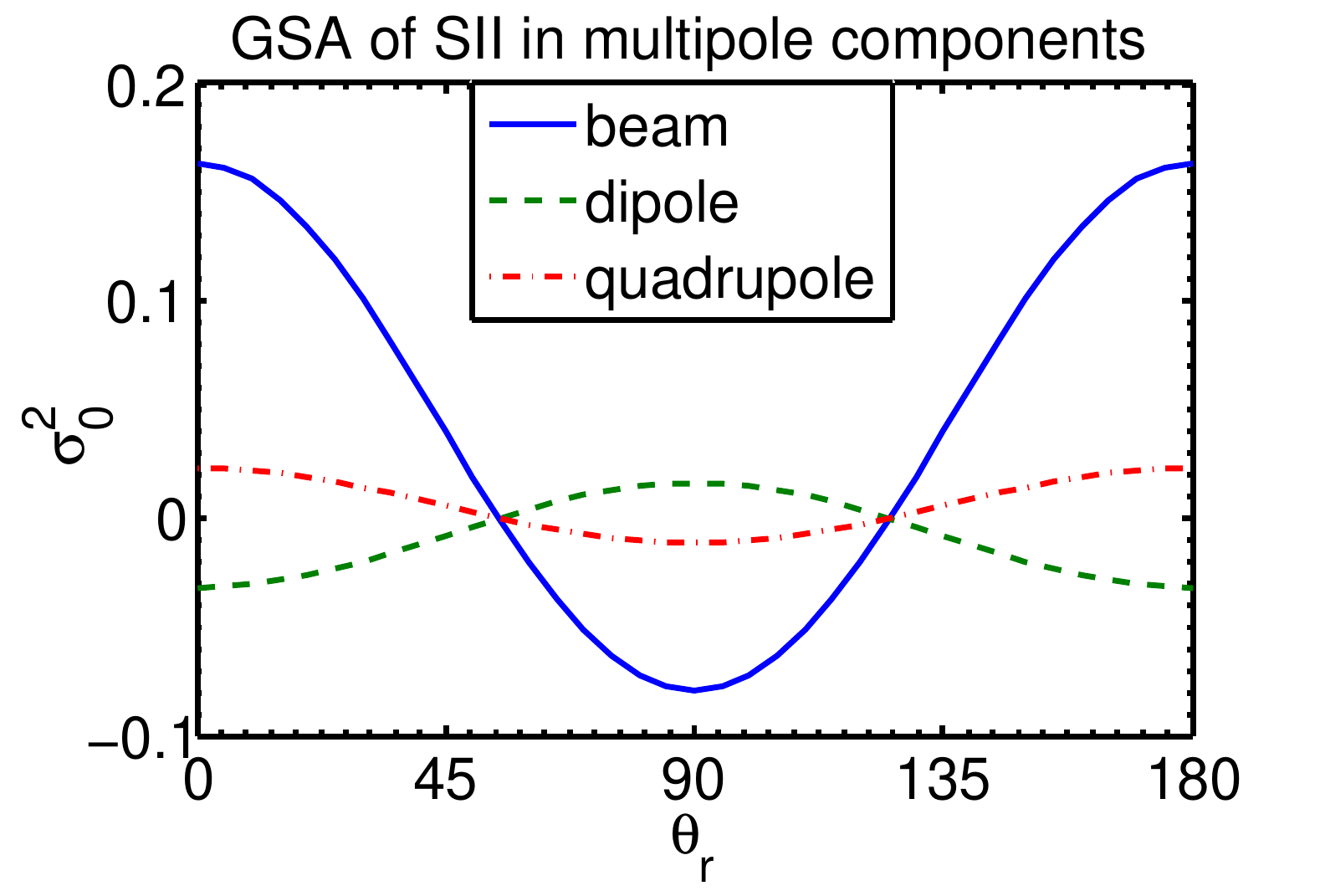}\label{fig14a}}
\subfigure[]{
\includegraphics[width=0.45\columnwidth,
 height=0.28\textheight]{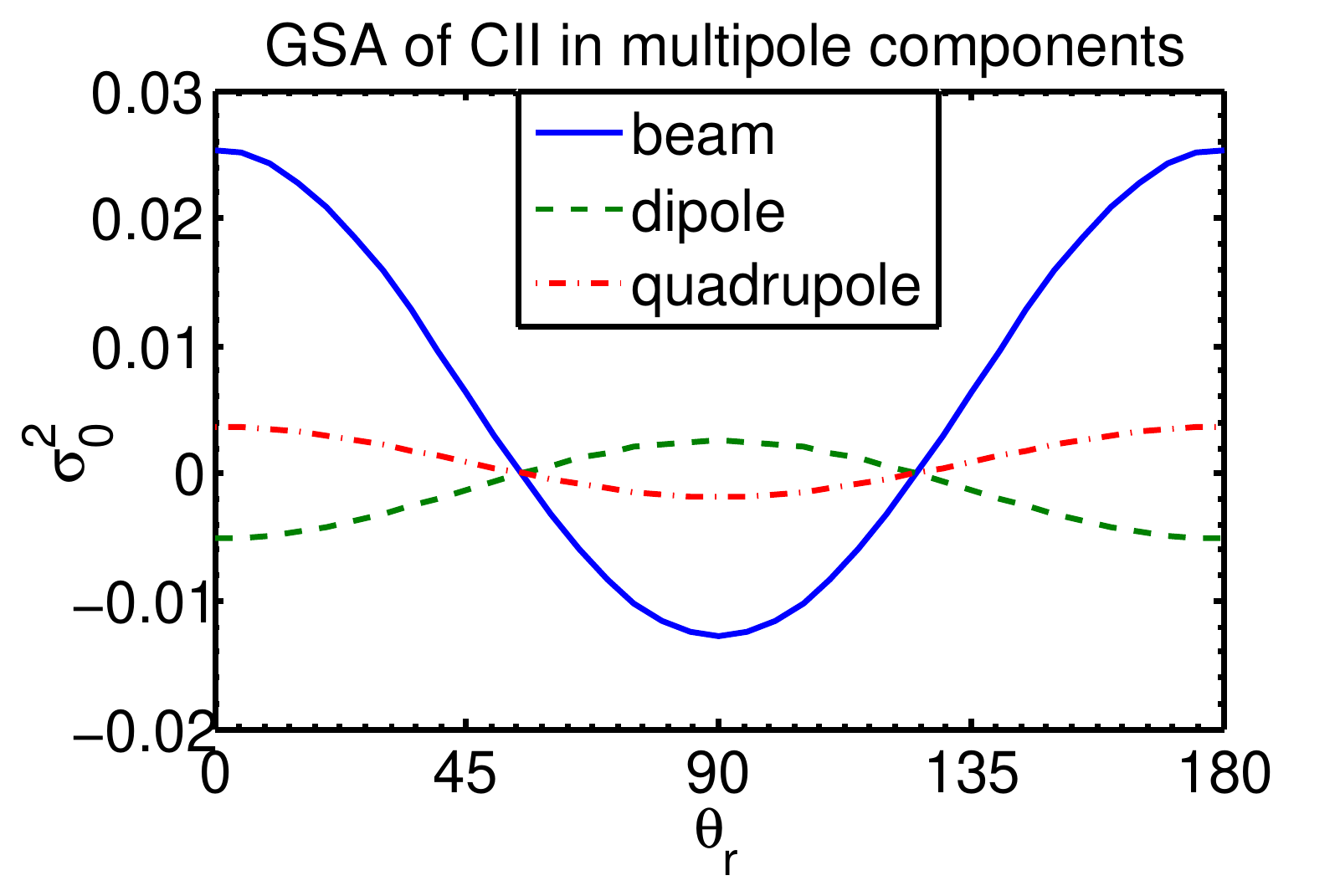}\label{fig14b}}
\caption{Comparison of the alignment parameter in a beam of light, dipole, and quadrupole radiation field vs. $\theta_r$ for (a) SII and (b) CII, respectively.
}
\end{figure}

\begin{figure}
\centering
\subfigure[]{
\includegraphics[width=0.32\columnwidth,
height=0.22\textheight]{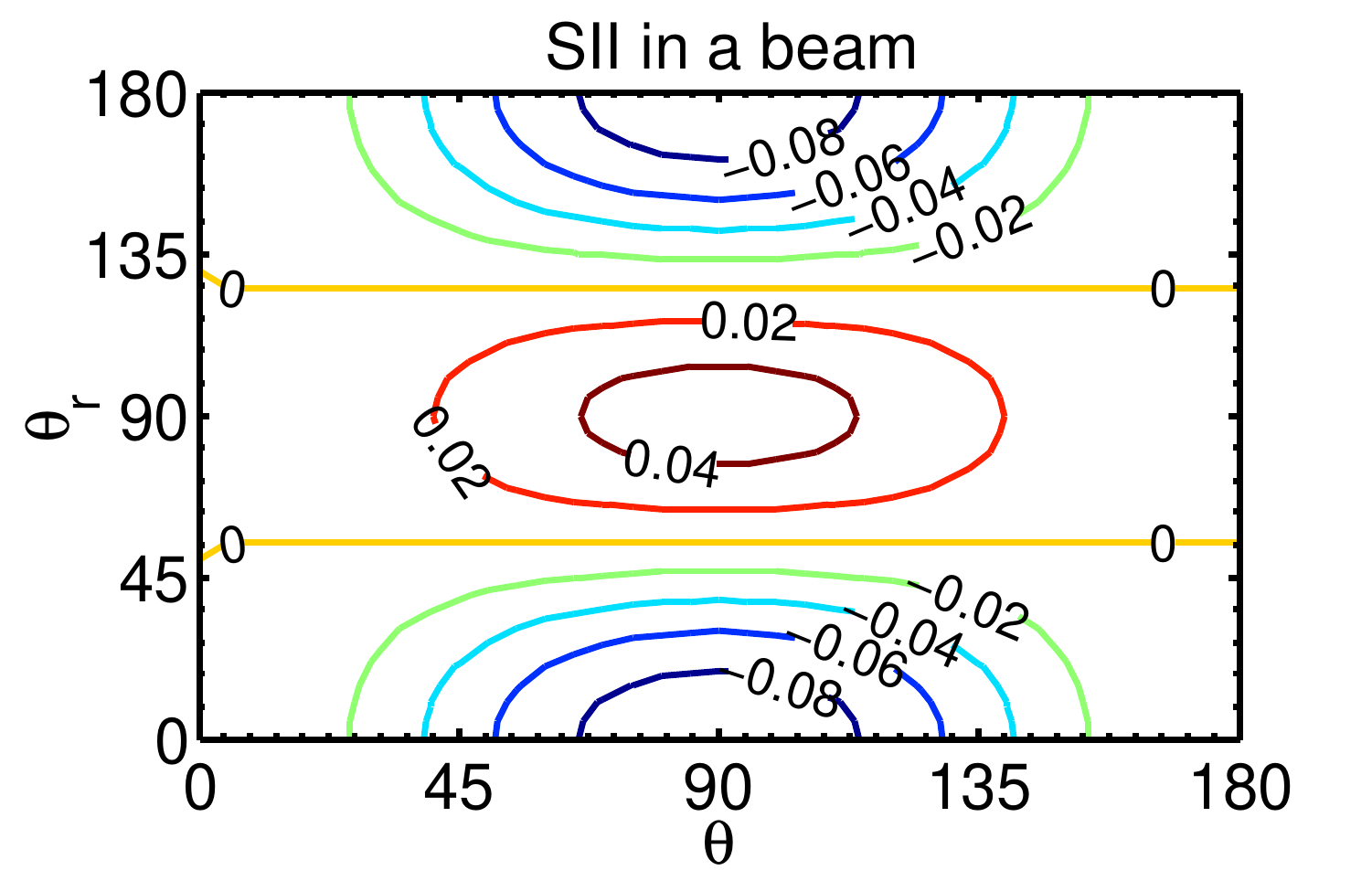}\label{fig15a}}
\subfigure[]{
\includegraphics[width=0.32\columnwidth,
height=0.22\textheight]{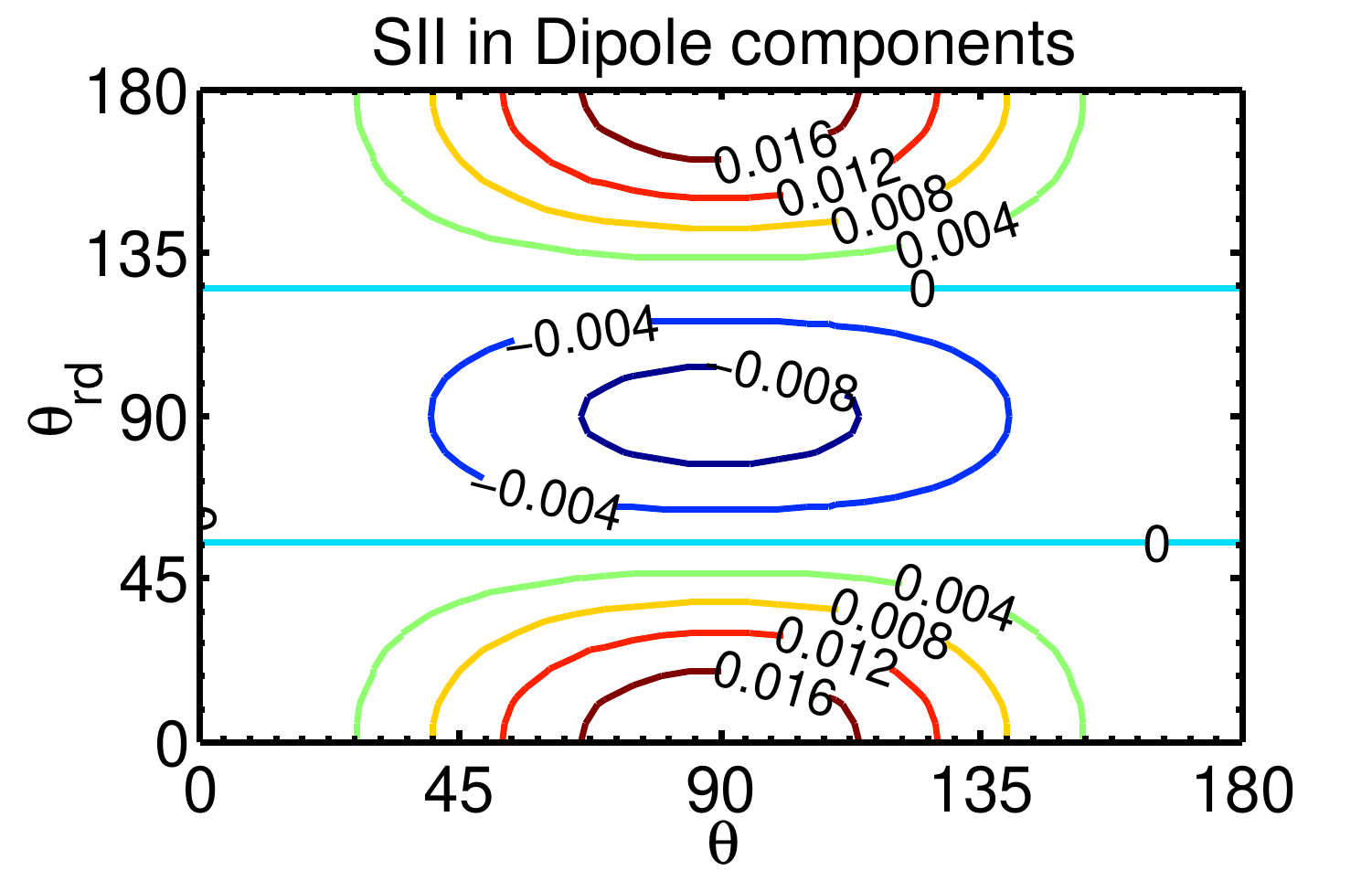}\label{fig15b}}
\subfigure[]{
\includegraphics[width=0.32\columnwidth,
height=0.22\textheight]{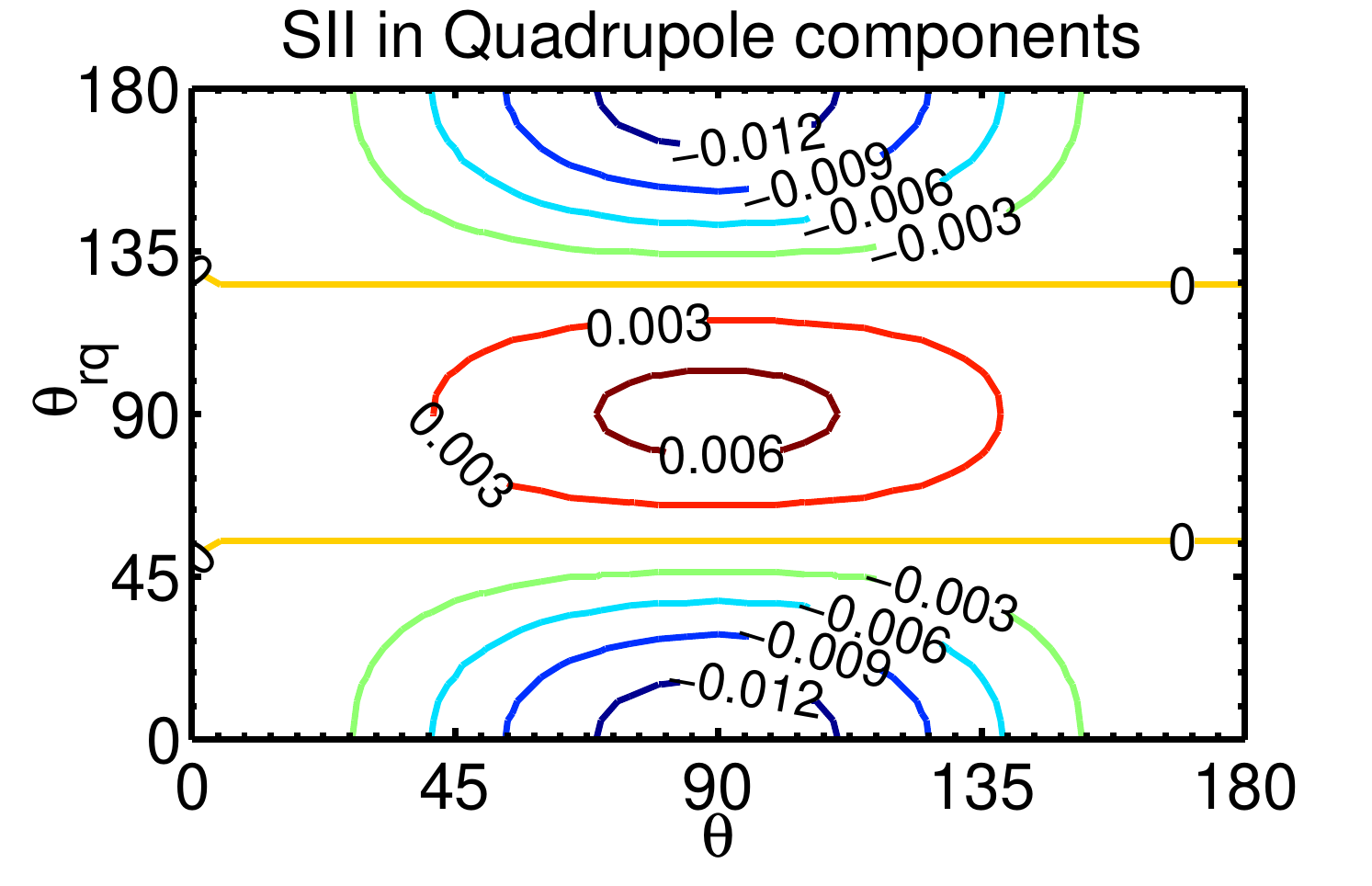}\label{fig15c}}
\caption{The iso-contour graphs for the polarization of SII absorption line. Different plots reveal the alignment induced by
(a) A beam of light;
(b) Dipole radiation field;
(c) Quadrupole radiation field, respectively.}
\end{figure}

\begin{figure}
\centering
\subfigure[]{
\includegraphics[width=0.32\columnwidth,
height=0.22\textheight]{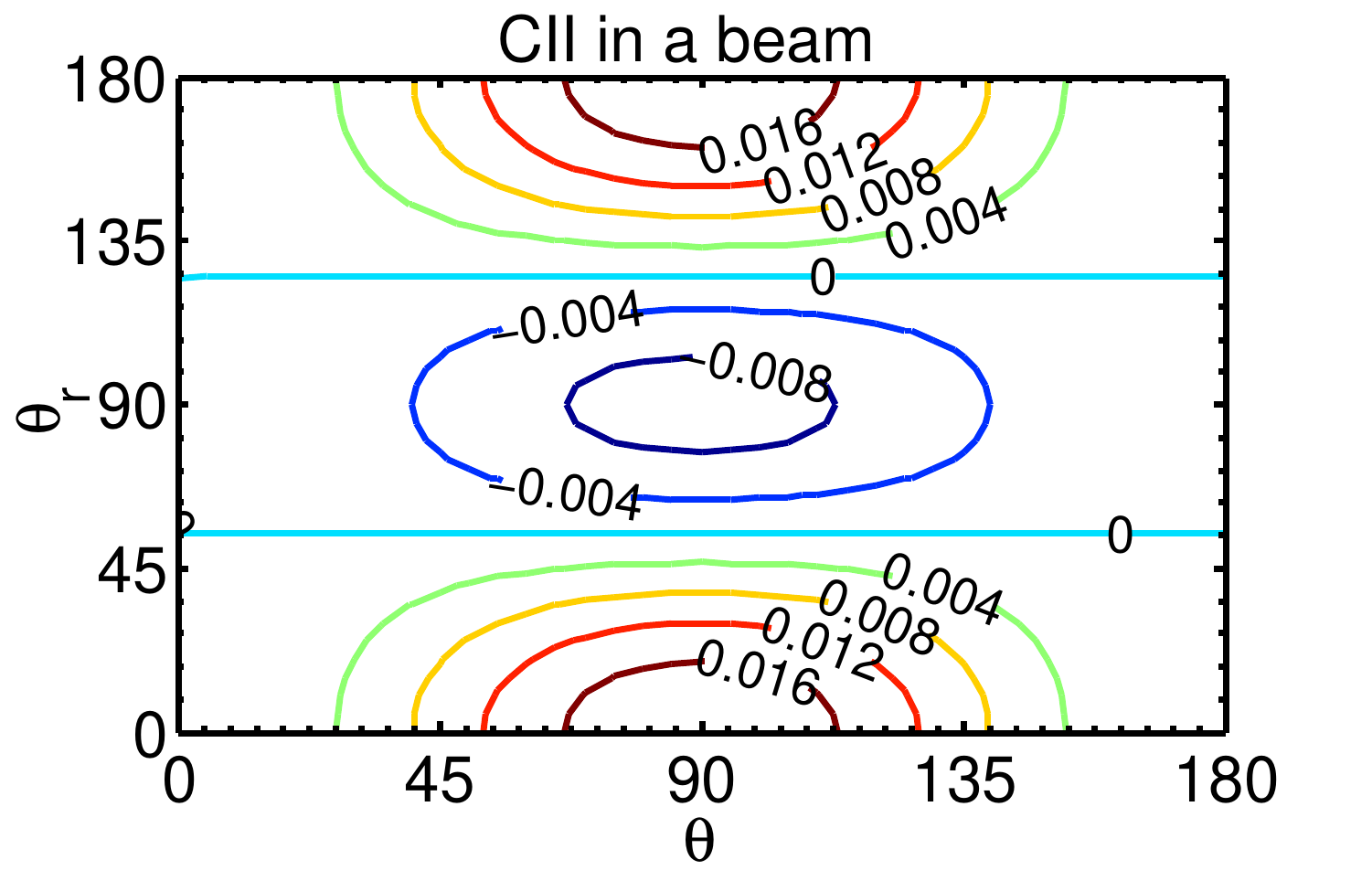}\label{fig16a}}
\subfigure[]{
\includegraphics[width=0.32\columnwidth,
height=0.22\textheight]{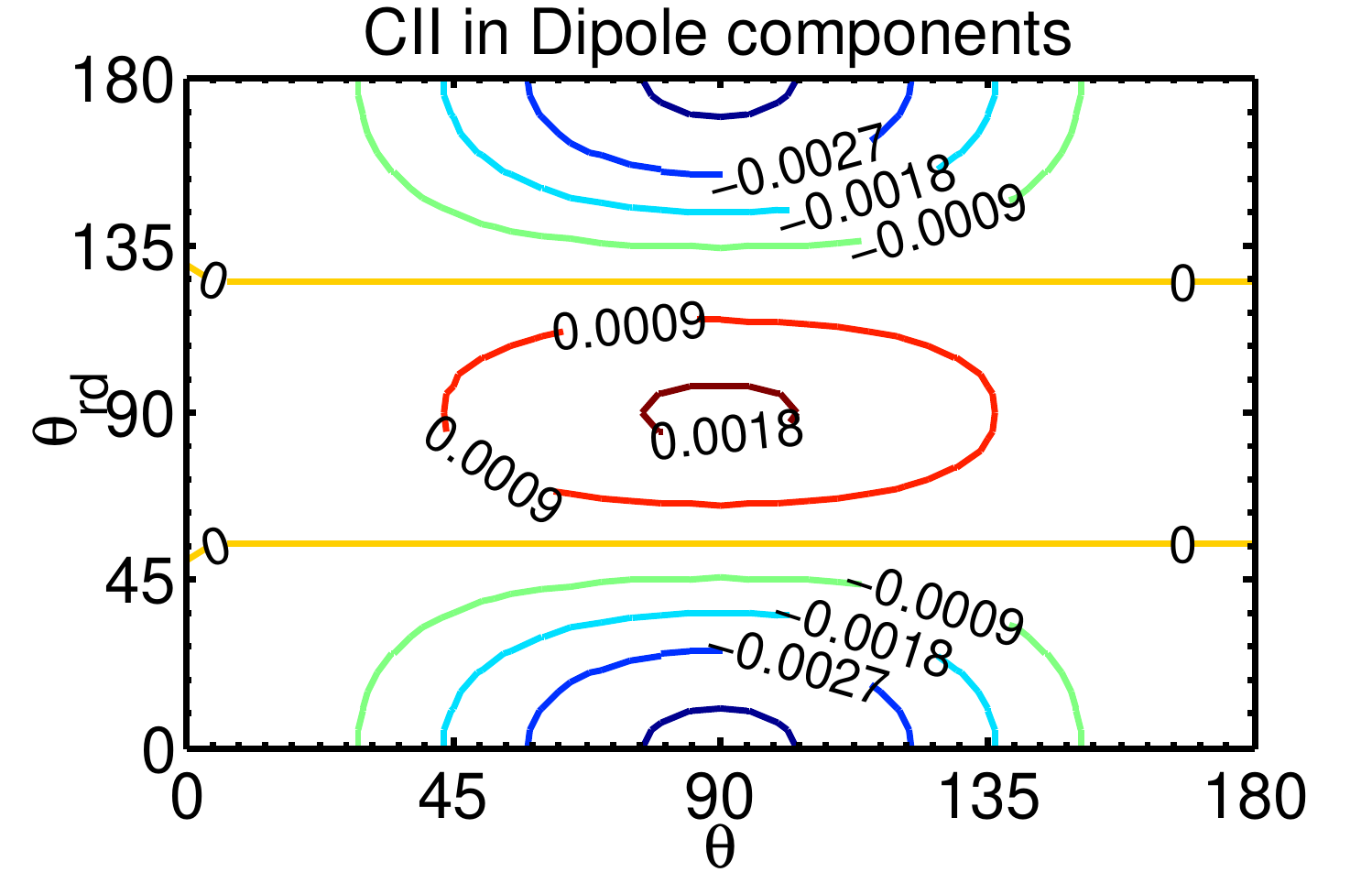}\label{fig16b}}
\subfigure[]{
\includegraphics[width=0.32\columnwidth,
height=0.22\textheight]{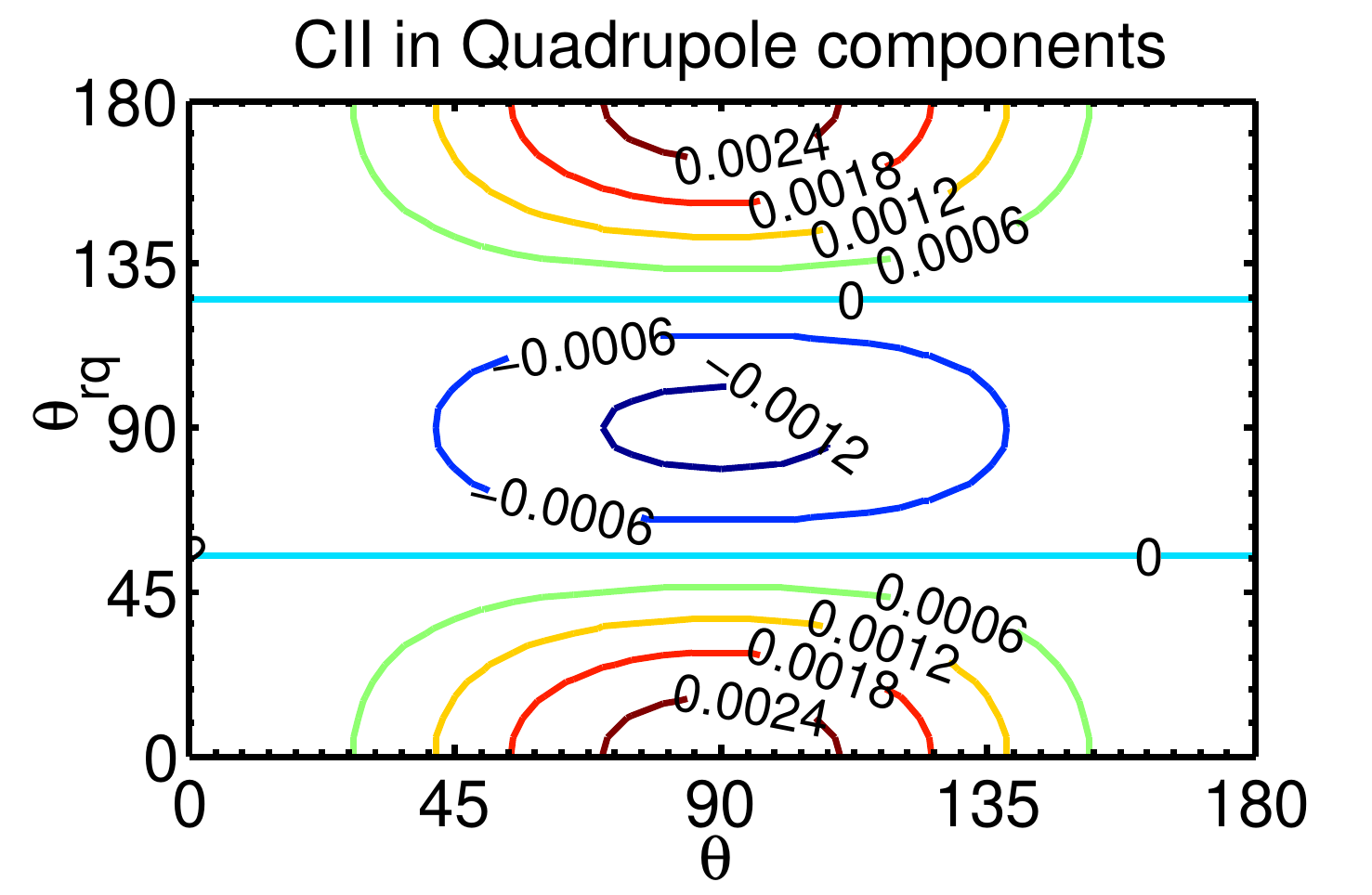}\label{fig16c}}
\caption{The iso-contour graphs for the polarization of CII absorption line. Different plots reveal the alignment induced by
(a) A beam of light;
(b) Dipole radiation field;
(c) Quadrupole radiation field, respectively.}
\end{figure}

Any radiation field can be decomposed in terms of irreducible representations of the rotational symmetry group, which can be represented by spherical harmonics and related sets of orthogonal functions $f(\theta,\phi) = \sum_{l=0}^\infty\, \sum_{m=-l}^{l}\, C^m_l\, Y^m_l(\theta,\phi)$ \citep[see][]{1975clel.book.....J}. The function $f(\theta,\phi)$ is the spatial distribution of the radiation ($\theta$, $\phi$). $Y^m_l(\theta,\phi)$ are the standard spherical harmonics, and $C^m_l$ are the coefficients. The quantity $l$ in both terms refers to the order of the multipole component. For example, $l=2$ denotes the dipole component, the intensity of which is
\begin{equation}\label{dipoledef}
\sum_{m=-2}^{2} C^m_2\, Y^m_2(\theta,\phi) \propto \sin^{2}\theta,
\end{equation}
and $l=4$ represents quadrupole components with the intensity
\begin{equation}\label{quaddef}
\sum_{m=-4}^{4} C^m_4\, Y^m_4(\theta,\phi) \propto \cos^{2}\theta\sin^{2}\theta.
\end{equation}
The overall alignment in the total radiation field could be obtained by studying the alignment in each multipole radiation component. As examples, GSA in dipole and quadrupole radiation field are discussed. As we shall demonstrate below, these are the dominant contributions to the alignment.

\subsection{Physics for GSA in multipole radiation field}

The dipole radiation field is presented in Fig.~\ref{fig12a} and Fig.~\ref{fig12b} in the direction perpendicular and parallel to the symmetrical axis, respectively. We define the coordinate system in Fig.~\ref{fig12c} and illustrate the coordinate transform needed to simplify the computation in Appendix~\ref{coordinatestransfer}. The radiation tensor can be obtained by inserting Eq.~\eqref{dipoledef} into Eq.~\eqref{tensordef}:
\begin{equation}
\bar{J}_{0}^{2}=\frac{1}{4\pi}\int_{0}^{2\pi}\int_{0}^{\pi}\frac{(1.5(\sin\theta_{rd}\sin\Psi_{kd}\cos\phi_{kd}+\cos\theta_{rd}\cos\Psi_{kd})-0.5)\sin^{3}\Psi_{kd}}{\sqrt{6}}\ud\Psi_{kd}\ud\phi_{kd}
\end{equation}

The irreducible radiation tensor in dipole radiation field is
\begin{equation}\label{dipoletensor}
\bar{J}_{0}^{2}=\frac{\sqrt6(1-3\cos^2\theta_{rd})}{90}.
\end{equation}

Infixing Eq.~\eqref{dipoletensor} into Eq.~\eqref{groundoccu}, we obtain the alignment parameter of SII in dipole radiation radiation field:
\begin{equation}\label{dipoledens}
\sigma_{0}^{2}(J_l)=\frac{1.6795\cos^{2}\theta_{rd}-0.5598}{34.8481-0.1986\cos^{2}\theta_{rd}}.
\end{equation}

By substituting Eq.~\eqref{dipoledens} into Eq.~\eqref{ptau}, the polarization of SII absorption line in dipole radiation radiation field is obtained:
\begin{equation}
\frac{P}{\tau}=\frac{(2.5193\cos^{2}\theta_{rd}-0.8397)\sin^{2}\theta \omega_{J_lJ_u}^2}{49.2827-0.2809\cos^{2}\theta_{rd}+(1.6795\cos^{2}\theta_{rd}-0.5598)(1-1.5\sin^{2}\theta)\omega_{J_lJ_u}^2}.
\end{equation}

The geometric distribution of quadrupole radiation field is presented in Fig.~\ref{fig13a} and Fig.~\ref{fig13b} in the direction perpendicular and parallel to the symmetrical axis, respectively. The coordinate system is defined in Fig.~\ref{fig13c}. We apply the similar procedure to calculate GSA in quadrupole radiation field:
\begin{equation}\label{quadtensor}
\bar{J}_{0}^{2}=\frac{\sqrt6(3\cos^2\theta_{rq}-1)}{630}
\end{equation}

The alignment parameter in quadrupole radiation field:
\begin{equation}\label{quaddens}
\sigma_{0}^{2}(J_l)=\frac{-0.2399\cos^{2}\theta_{rq}+0.0800}{6.9923-0.0284\cos^{2}\theta_{rq}},
\end{equation}
and the polarization in quadrupole radiation field:
\begin{equation}
\frac{P}{\tau}=\frac{(-0.3599\cos^{2}\theta_{rq}+0.1200)\sin^{2}\theta \omega_{J_lJ_u}^2}{9.8886-0.0402\cos^{2}\theta_{rq}+(-0.2399\cos^{2}\theta_{rq}+0.0800)(1-1.5\sin^{2}\theta)\omega_{J_lJ_u}^2}.
\end{equation}

$\bar{J}_{0}^{2}$ is equal to zero at the Van Vleck angle $\theta_{rd}=\theta_{rq}=54.7^{\circ}$ for both dipole and quadrupole radiation field. As a result, the polarization flips between being parallel and perpendicular to the direction of magnetic field in both the dipole and quadrupole radiation field at the Van Vleck angle.

\subsection{Applications of the multipole expansion in the intersteller medium}

\subsubsection{Alignment in mutipole radiation field}

The alignment in different multipole radiation fields are compared with that in a beam of light in Fig. 14. Fig. 15, 16 respectively present the polarization of SII and CII absorption lines in a beam of light, dipole radiation field, and quadrupole radiation field. The alignment is reduced in the dipole radiation field and even further in quadrupole radiation field compared to that in a beam of light since the dipole and quadrupole radiation field are more isotropic. The results suggest that even higher order of radiation components can be neglected and a good approximation can be made by counting radiation components up to the quadrupole component. In addition, the observed polarization reaches a maximum at $\theta=\pi/2$ with the same $\theta_r$ (see also Eq.\ref{ptau}). The switch of the polarization between being parallel and perpendicular to magnetic field in dipole and quadrupole radiation field happens at the Van Vleck angle, same as the case in a beam of light, regardless of the atomic species. As a result, {\em the 2D information of magnetic field can be directly obtained by the direction of polarization with a $90^\circ$ degeneracy. If the degree of polarization is observed, 3D magnetic field can be obtained.}

\subsubsection{Polarization in multipole components}

The aim of the multipole expansion is to approximate the main function by taking only first few components into account. An sufficient approximation of radiation field can be made by using dipole and quadrupole radiation component since higher order radiation fields are isotropic to such a degree that the resulting alignment is very weak. Therefore, we shall only discuss GSA in dipole and quadrupole radiation field here.

\begin{figure}
\centering
\subfigure[]{
\includegraphics[width=0.45\columnwidth,
 height=0.28\textheight]{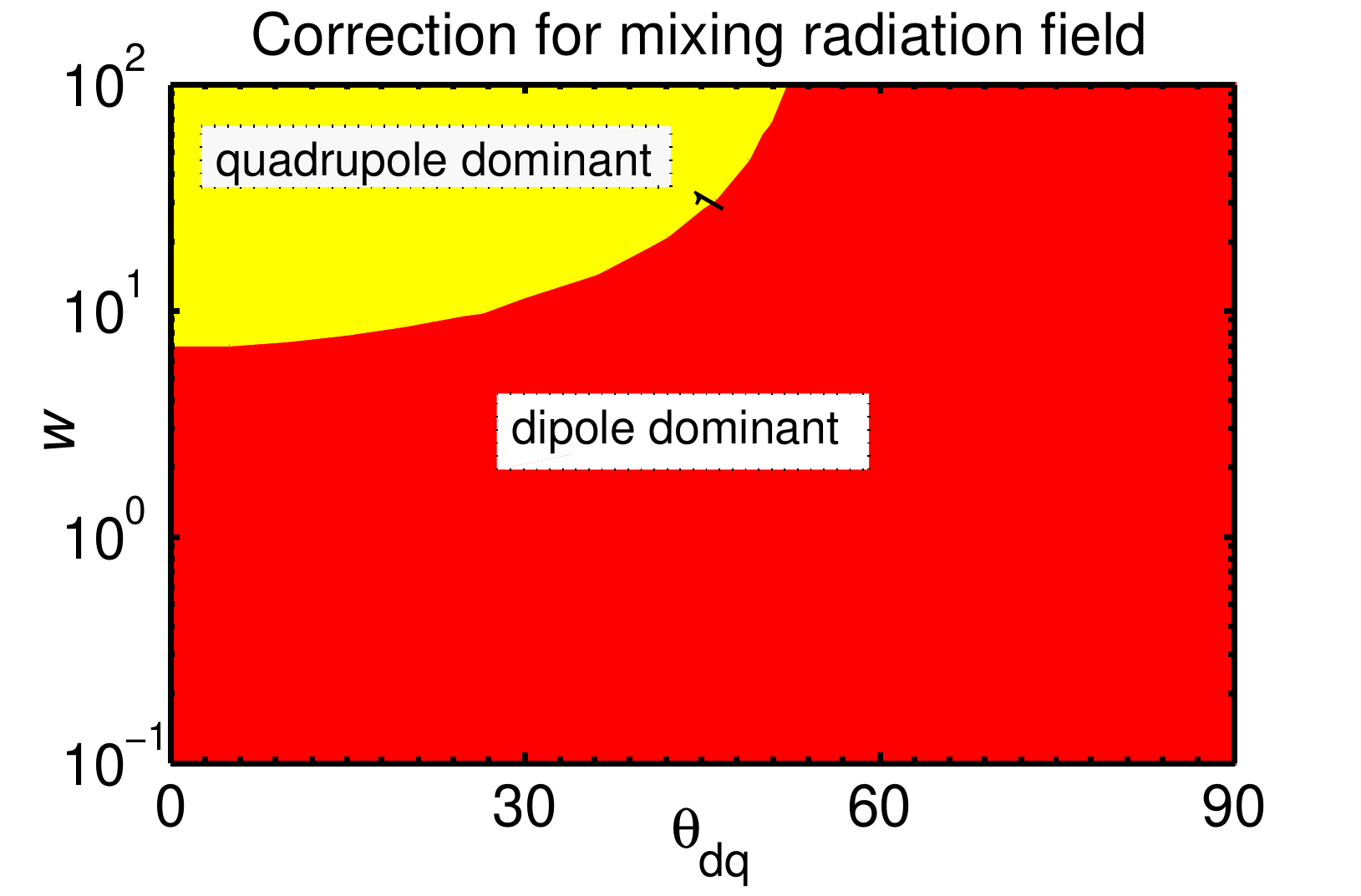}\label{fig17a}}
\subfigure[]{
\includegraphics[width=0.45\columnwidth,
 height=0.28\textheight]{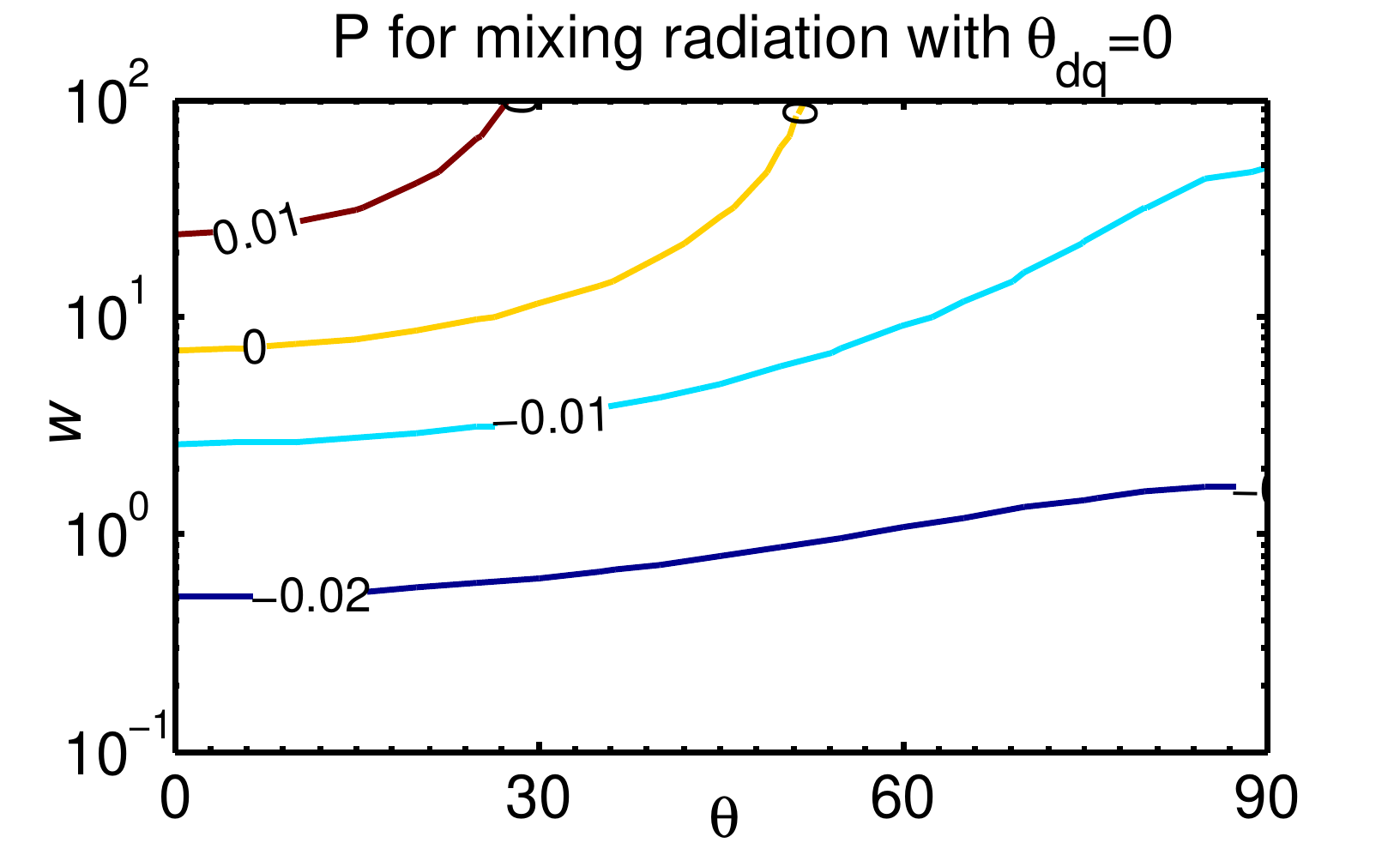}\label{fig17b}}
\caption{Polarization induced by GSA with multipole radiation field.
(a) Iso-contour plot for the correction of the polarization $R_P$ for SII absorption line from adding quadrupole radiation component to a dipole radiation field. $\theta_{dq}$: the angle between dipole and quadrupole axis; $w$: the intensity ratio of quadrupole radiation field to dipole radiation field. The division line is where the contributions from the dipole radiation field and quadrupole radiation field cancel each other. The regions where dipole dominant region and quadrupole dominates are noted.
(b) Iso-contour plot for the polarization of SII absorption line in the case when the dipole axis coincides with the quadrupole axis. $\theta$: the angle between line of sight and magnetic field.
}
\end{figure}

We add quadrupole radiation field to the dipole field with varying intensity ratio ($w$) and angle between their axes ($\theta_{dq}$). The intensity ratio of quadrupole radiaition field and dipole radiation is defined as:
\begin{equation}
w\equiv u_q/u_d,
\end{equation}
where the quantities $u_d$ and $u_q$ are the intensity averaged over the whole solid angle for dipole and quadrupole radiation field, respectively. The correction due to the addition of the quadrupole field is defined as:
\begin{equation}
R_{P}\equiv \frac{\Delta P}{P}=1-\frac{P(dipole)}{P(mix)}.
\end{equation}
Fig.~\ref{fig17a} presents the correction of the polarization $R_P$ with different intensity ratio ($w$) and different angle between their axes ($\theta_{dq}$). Evidently, the anisotropy of the dipole and the quadrupole radiation fields are opposite. The division line marks the place where the contributions from the dipole radiation field and the quadrupole radiation field cancel each other. It is clear that the contribution from dipole radiation field is dominant over that from quadrupole radiation field if $w<7$. {\em This is a general feature for radiation field.} As an example, Fig.~\ref{fig17b} is plotted to illustrate the polarization of SII absorption line induced by GSA when the dipole axis coincides with the quadrupole axis.

\section{Discussions}

In the paper, we extend the studies of GSA to general radiation field in diffuse medium. It is concluded in the paper that GSA can be applied to trace magnetic field in diffuse medium in general radiation field, which largely broadens the applicability of GSA in astrophysical circumstances. Calculations for both absorption and emission lines are presented, filling in the gap for earlier studies on GSA by considering the atomic alignment in different radiation field. In addition, we adopt the method of multipole expansion in order to obtain the polarization in general radiation field with unidentified pumping sources.

{\em Study of GSA for emission line}

We have discussed the results of absorption line in extended radiation field. We stress that the emission line can also be used to trace magnetic field in all the situations discussed in the paper. As an illustration, we present the calculation for emission lines in circumstellar medium. Similar approaches can be applied to study the polarization of emission lines in other circumstances.

{\em Tracing the 3D magnetic field with GSA}

It is concluded that the direction of polarization provide us with the projection of magnetic field on the plane of sky with a $90^\circ$ degeneracy in general radiation field. Results on switch angles for different radiation field are provided in Table 1. 3D magnetic field could be obtained if the degree of polarization is observed.

\begin{table}[!hbp]\label{turnoff}
\centering
\caption{Switch angles of the polarization induced by GSA.}
\begin{tabular}{|c|c|c|}
\hline
Diffuse medium & Switch angle & Notes \\
\hline
Circumstellar medium & $54.7^{\circ}$ & the Van Vleck angle \\
Binary systems & $\arccos\frac{1}{\sqrt{3t}}$ & t is defined in Eq.~\eqref{tdef} \\
Disc shape radiation field & $\arccos\sqrt{\frac{2}{3f_l(\alpha_0)}}$ & $f_l(\alpha_0)$ is defined in Eq.~\eqref{fldef} \\
the Local ISM & $35.3^{\circ}$ & Example for Disc shape radiation field \\
Dipole radiation field & $54.7^{\circ}$ & the Van Vleck angle \\
Quadrupole radiation field & $54.7^{\circ}$ & the Van Vleck angle \\
\hline
\end{tabular}
\end{table}

{\em Atomic alignment in identified radiation sources}

We have provided a tool to trace magnetic field in circumstellar medium, binary systems, disc, and the Local ISM. Apparently, similar approaches can be applied to study the alignment in diffuse medium near other radiation sources. For instance, multiple systems can be studied by considering each star as a point source and integrating all the incoming lights to the medium.

{\em Strong pumping regime}

The strong pumping approximation is specifically discussed for the alignment in circumstellar medium in the paper. Evidently, this approximation is also applicable for the alignment in similar systems where the medium is sufficiently close to the sources. The criterial is given by Eq.~\eqref{radiuscal}.

{\em The influence of collision}

We discuss the diffuse medium where the influence of collision is negligible. Collisions redistribute atoms to different sublevels though with reduced efficiency as disalignment of the ground state requires spin flips (see \citealt{Hawkins:1955fv}). We shall evaluate the effect of collision when it cannot be neglected on GSA induced by extended radiation field in future work.

\section{Summary}

GSA with different radiation fields are discussed in the study. GSA has been demonstrated in earlier studies a powerful magnetic tracer when the radiation field is a beam of light. We extend the applicability of GSA to the medium in extended radiation fields including circumstellar medium, binary systems, disc, and the Local Interstellar Medium. Moreover, the method of multipole expansion is utilized for the radiation field with unidentified pumping sources. It is concluded that:

\begin{enumerate}

\item GSA exists wherever the radiation field are anisotropic.
\item Spectropolarimetry modulated by GSA can be applied to trace magnetic field in diffuse medium in general radiation field.
\item The direction of polarization directly traces 2D magnetic field in the plane of sky. Same as the case in a beam of light, the polarization of absorption line is either parallel or perpendicular to magnetic field in extended radiation field. We provide the criteria at which the polarization flips for all the cases.
\item Three-dimension (3D) mapping of magnetic field can be determined with quantitative measurement of the polarization from GSA.
\item The alignment with unidentified radiation sources can be obtained by considering the alignment in the dipole and quadrupole radiation component of the radiation field.
\end{enumerate}

\begin{acknowledgements}
We have benefited from valuable discussions with Thijs Kouwenhoven, Jinyi Shangguan and Qingjuan Zhu. The support by the Templeton senior grant from Beyond the Horizon program is acknowledged.
\end{acknowledgements}

\begin{appendix}
\section{A. FROM MAGNETIC FRAME TO RADIATION FRAME}\label{coordinatestransfer}

The symmetry axis of the radiation may not be parallel to magnetic field in circumstellar medium. Hence, a transformation from the magnetic frame to the extended radiation frame is needed to integrate the influence of the whole cone to magnetic field. We illustrate the geometry for the coordinate conversion in Fig.~\ref{fig2c}. We use ($\theta_{o}$, $\phi_{o}$) for the angle coordinates in extended frame ($x''y''z''-$frame), ($\theta_{r}$, $\phi_{r}$) for the angle coordinates in the magnetic frame. ($\theta_B,$ $\phi_B$) are the coordinates for the axis of the radiation cone in $xyz-$frame in Fig.~\ref{fig2b}. We consider a line whose length equals to $1$ in the magnetic frame with the coordinates:
\begin{equation}
\left(
\begin{array}{ccc}
\sin\theta_{r}\cos\phi_{r} & \sin\theta_{r}\sin\phi_{r} & \cos\theta_{r}
\end{array}
\right),
\end{equation}
whereas in the extended radiation frame the coordinates are
\begin{equation}
\left(
\begin{array}{ccc}
\sin\theta_{o}\cos\phi_{o} & \sin\theta_{o}\sin\phi_{o} & \cos\theta_{o}
\end{array}
\right).
\end{equation}

The relation between these two coordinates can be demonstrated by
\begin{equation}
\begin{split}
\left(
\begin{array}{ccc}
\cos\theta_{r}\cos\phi_{r} & \sin\theta_{r}\cos\phi_{r} & \sin\phi_{r} \\
\end{array}
\right)=&
\left(
\begin{array}{ccc}
\cos\theta_{o}\cos\phi_{o} & \sin\theta_{o}\cos\phi_{o} & \sin\phi_{o} \\
\end{array}
\right)\\
&\left(
\begin{array}{ccc}
\sin\phi_B & \cos\phi_B & 0 \\
-\cos\phi_B & \sin\phi_B & 0 \\
0 & 0 & 1 \\
\end{array}
\right)
\left(
\begin{array}{ccc}
1 & 0 & 0 \\
0 & \cos\theta_{B} & \sin\theta_{B} \\
0 & -\sin\theta_{B} & \cos\theta_{B} \\
\end{array}
\right).
\end{split}
\end{equation}

Thus for emission line,
\begin{equation}
\theta_{r}=\arccos\left(\cos\phi_{B}\sin\theta_{B}\sin\theta_{o}\cos\phi_{o}+\sin\phi_{B}\sin\theta_{B}\sin\theta_{o}\sin\phi_{o}+\cos\theta_{B}\cos\theta_{o}\right),
\end{equation}
\begin{equation}
\phi_{r}=\arctan\left(\frac{\sin\phi_{B}\cos\theta_{B}\sin\theta_{o}\cos\phi_{o}+\cos\phi_{B}\cos\theta_{B}\sin\theta_{o}\sin\phi_{o}-\sin\theta_{B}\cos\theta_{o}}{\sin\phi_{B}\sin\theta_{o}\cos\phi_{o}-\cos\phi_{B}\sin\theta_{o}\sin\phi_{o}}\right).
\end{equation}

For absorption line, we only need to consider $\sigma_0^2$ (see \S 2.2 for details). Therefore, we set
\begin{equation}
\sin\phi_{B}=1.
\end{equation}

And the angle coordinates in extended frame ($x''y''z''-$frame) are
\begin{equation}
\theta_{r}=\arccos\left(\sin\theta_{B}\sin\theta_{o}\sin\phi_{o}+\cos\theta_{B}\cos\theta_{o}\right),
\end{equation}
\begin{equation}
\phi_{r}=\arctan\left(\frac{\sin\theta_{o}\sin\phi_{o}\cos\theta_{B}-\cos\theta_{o}\sin\theta_{B}}{\sin\theta_{o}\cos\phi_{o}}\right).
\end{equation}

\section{B. SUMMARY OF ANGLES USED IN THE PAPER}

\begin{longtable}{|l|c|}
\hline
angle between line of sight and magnetic field in Fig.~\ref{fig1a} & $\theta$ \\
azimuth coordinates in axial frame where line of sight is $z-$axis & $\phi$ \\
cone angle for radiation in circumstellar medium in Fig.~\ref{fig2a} & $\theta_c$ \\
angle between radiation direction and magnetic field in Fig.~\ref{fig2b} & $\theta_r$ \\
azimuth coordinates in $xyz-$frame in Fig.~\ref{fig2b} & $\phi_r$ \\
angle between cone axis and magnetic field in Fig.~\ref{fig2b} & $\theta_B$ \\
azimuth angle for cone axis in $xyz-$frame in Fig.~\ref{fig2b} & $\phi_B$ \\
angle between radiation and cone axis in Fig.~\ref{fig2c} & $\theta_o$ \\
azimuth coordinates in $x''y''z''-$frame in Fig.~\ref{fig2c} & $\phi_o$ \\
angle between magnetic field and the plane of radiation in binary in Fig.~\ref{fig6b} & $\phi_b$ \\
angles in the plane of radiation in Fig.~\ref{fig6b} & $\alpha_1,\alpha_2$ \\
angle between two radiation line in binary in Fig.~\ref{fig6b} & $\psi$ \\
angle between magnetic field and the plane of pumping disc in Fig.~\ref{fig9a} & $\phi_l$ \\
flare angle of disc shape radiation field in Fig.~\ref{fig9a} & $\alpha_0$ \\
angle between dipole axis and magnetic field in Fig.~\ref{fig12c} & $\theta_{rd}$ \\
angle between radiation direction and dipole axis in Fig.~\ref{fig12c} & $\Psi_{kd}$ \\
azimuth coordinates in dipole frame in Fig.~\ref{fig12c} & $\phi_{kd}$ \\
angle between quadrupole axis and magnetic field in Fig.~\ref{fig13c} & $\theta_{rq}$ \\
angle between radiation direction and quadrupole axis in Fig.~\ref{fig13c} & $\Psi_{kq}$ \\
azimuth coordinates in quadrupole frame in Fig.~\ref{fig13c} & $\phi_{kq}$ \\
angle between dipole axis and quadrupole axis & $\theta_{dq}$ \\
\hline
\end{longtable}\label{angles}
\section{C. DEFINITION OF SOME PHYSICAL PARAMETERS}\label{parametersforeq}
The basic formulae for GSA have been well illustrated in earlier studies\citep[see, e.g.,][]{YLfine,YLhyf,YLhanle}. Only some key terms are presented here for the sake of the complicity of the paper.

The Einstein coefficients illustrating the stimulated transitions $B$ are related to the Einstein spontaneous emission rate $A$ by:
\begin{equation}\label{einstein}
[J_u]A(J_u\rightarrow J_l)=\frac{2h\nu^3}{c^2}[J_l]B_{lu}=\frac{2h\nu^3}{c^2}[J_u]B(J_u\rightarrow J_l)
\end{equation}
The absorption coefficients $\eta_i$ to illustrate the polarized absorption from the ground level are defined as (\citealt{Landi-DeglInnocenti:1984kl}, see also \citealt{YLfine}):
\begin{equation}\label{etai}
\eta_i(\nu,\Omega)=\frac{h\nu_0}{4\pi}Bn(J_l)\Psi(\nu-\nu_0)\sum_{\substack{KQ}}(-1)^K\omega^{K}_{J_lJ_u}\sigma^K_Q(J_l)\mathcal{J}^K_Q(i,\Omega),
\end{equation}
where the total atomic population $n(J_l)$ on the lower level $J_l$ is defined as $n\sqrt{[J_L]}\rho^0_0(J_l)$. Additionally,
\begin{equation}\label{omegai}
\omega_i\equiv\left\{\begin{array}{ccc}1&1&K\\J_l&J_l&J_u\end{array}\right\}/\left\{\begin{array}{ccc}1&1&0\\J_l&J_l&J_u\end{array}\right\}.
\end{equation}
Moreover, the emission coefficients $\epsilon_i$ demonstrating the polarized emission from the upper level are \citep[see][]{YLhyf}:
\begin{equation}\label{epsiloni}
\epsilon_i(\nu,\Omega)=\frac{h\nu_0}{4\pi}An(J_u,\theta_r)\Psi(\nu-\nu_0)\sum_{\substack{KQ}}\omega^{K}_{J_uJ_l}\sigma^K_Q(J_u,\theta_r)\mathcal{J}^K_Q(i,\Omega).
\end{equation}

\section{D. DENSITY MATRIX}\label{densitymat}
The irreducible tensorial formulae adopted in the paper can be represented by the standard density matrix of atoms:
\begin{equation}
\rho^K_Q(J,J')=\sum_{\substack{MM'}}(-1)^{J-M}(2K+1)^\frac{1}{2}\left(\begin{array}{ccc}J&K&J'\\-M&Q&M'\end{array}\right)<JM|\rho|J'M'>
\end{equation}
Additionally, the irreducible spherical tensor for photons is:
\begin{equation}
\bar{J}^K_Q(J,J')=\sum_{\substack{qq'}}(-1)^{1+q}[3(2K+1)]^\frac{1}{2}\left(\begin{array}{ccc}1&1&K\\q&-q'&-Q\end{array}\right)J_{qq'}
\end{equation}
\end{appendix}
\label{lastpage}
 \bibliographystyle{plainnat}
 \setcitestyle{square,aysep={},yysep={;}}
 \bibliography{Zhan0306}
\end{document}